\documentclass[aps,prc,twocolumn,superscriptaddress,letterpaper]{revtex4-1} 
\usepackage[final]{changes} \setremarkmarkup{(#2)} \colorlet{Changes@Color}{teal}
\definechangesauthor[name={Questions},color=red]{?}
\usepackage{titlesec}
\usepackage[titletoc,toc,title]{appendix}
\usepackage{enumerate}
\usepackage{float}
\usepackage{color}
\usepackage{comment}
\usepackage{siunitx}
\usepackage{soul}
\usepackage{graphicx}  
\usepackage{dcolumn}   
\usepackage{multirow}  
\usepackage{bm}        
\usepackage{amssymb}   
\usepackage{amsmath}   
\usepackage{epstopdf}
\usepackage{url}
\usepackage{hyperref}  
\usepackage{breakurl}
\hypersetup{breaklinks = true, colorlinks = true, citecolor = blue, linkcolor = blue, urlcolor = blue}
\usepackage{enumitem}
\usepackage{enumerate}
\usepackage{subfigure}
\definecolor{shadecolor}{RGB}{180,180,180}
\usepackage[most]{tcolorbox}
\usepackage{soul}
\usepackage{makecell}
\usepackage{listings}
\usepackage{csquotes}
\usepackage{graphicx,bm}
\usepackage{verbatim}
\usepackage[]{cleveref}

\crefname{section}{Sec.}{Secs.}
\Crefname{section}{Section}{Sections}
\crefname{figure}{Fig.}{Figs.}
\Crefname{figure}{Figure}{Figures}
\crefname{equation}{Eq.}{Eqs.}
\Crefname{equation}{Equation}{Equations}
\crefname{table}{Table}{Tables}
\Crefname{table}{Table}{Tables}

\titleformat{\section}{\bfseries\centering}{\large\Roman{section}.}{1em}{}
\titleformat{\subsection}{\bfseries\centering}{\Alph{subsection}.}{1em}{}

\newcommand{\subfigimg}[5][,]{%
  \setbox1=\hbox{\includegraphics[#1]{#3}}
  \leavevmode\rlap{\usebox1}
  \rlap{\hspace*{#5pt}\raisebox{\dimexpr\ht1-#4\baselineskip}{#2}}
  \phantom{\usebox1}
}

\usepackage{lineno}
\usepackage{blindtext}
\newcommand{\neue}{$\nu_{e}$}
\newcommand{\neumu}{$\nu_{\mu}$}

\newcommand{\uB}{MicroBooNE}

\newcommand{\CCQEp}{${\nu_\mu n \rightarrow \mu^{-} p}$~}

\newcommand{\MINERvA}{MINER$\nu$A}

\newcommand{\mup}{$\mu p$}
\newcommand{\CCIpOpi}{CC$1p0\pi$}

\newcommand{\RdQ}{$R_{\Delta Q}$}

\newcommand{\dEdx}{$dE/dx$}

\newcommand{\ChiSqrP}{$\chi^{2}_{\textrm{p}}$}

\newcommand{\CutChiSqrP}{ $80<(\chi^{2}_{\textrm{p}})^{\mu}$ and $(\chi^{2}_{\textrm{p}})^{p}<30$}
\newcommand{\Cutlmup}{ $l_{\mu}>l_{p}$}
\newcommand{\CutCollinearity}{ $| \theta_{12}- 90^{\circ}| \le 55^{\circ}$}
\newcommand{\CutDeltaPhi}{ $| \Delta \phi - 180^{\circ}| \le 35^{\circ}$}
\newcommand{\CutPt}{ $p_T \le 0.35$ GeV/c}

\newcommand{\CutFlashYZ}{$d_{\textrm{YZ}} < 200$ cm}
\newcommand{\CutFlashPE}{$N_{PE} > 150$}
\begin{document}
	\title{\huge{Rejecting cosmic background for exclusive neutrino interaction studies with Liquid Argon TPCs; a case study with the MicroBooNE detector}}%
	
\newcommand{\Bern}{Universit{\"a}t Bern, Bern CH-3012, Switzerland}
\newcommand{\BNL}{Brookhaven National Laboratory (BNL), Upton, NY, 11973, USA}
\newcommand{\Cambridge}{University of Cambridge, Cambridge CB3 0HE, United Kingdom}
\newcommand{\Chicago}{University of Chicago, Chicago, IL, 60637, USA}
\newcommand{\Cincinnati}{University of Cincinnati, Cincinnati, OH, 45221, USA}
\newcommand{\CSU}{Colorado State University, Fort Collins, CO, 80523, USA}
\newcommand{\Columbia}{Columbia University, New York, NY, 10027, USA}
\newcommand{\Davidson}{Davidson College, Davidson, NC, 28035, USA}
\newcommand{\FNAL}{Fermi National Accelerator Laboratory (FNAL), Batavia, IL 60510, USA}
\newcommand{\Harvard}{Harvard University, Cambridge, MA 02138, USA}
\newcommand{\IIT}{Illinois Institute of Technology (IIT), Chicago, IL 60616, USA}
\newcommand{\KSU}{Kansas State University (KSU), Manhattan, KS, 66506, USA}
\newcommand{\Lancaster}{Lancaster University, Lancaster LA1 4YW, United Kingdom}
\newcommand{\LANL}{Los Alamos National Laboratory (LANL), Los Alamos, NM, 87545, USA}
\newcommand{\Manchester}{The University of Manchester, Manchester M13 9PL, United Kingdom}
\newcommand{\MIT}{Massachusetts Institute of Technology (MIT), Cambridge, MA, 02139, USA}
\newcommand{\Michigan}{University of Michigan, Ann Arbor, MI, 48109, USA}
\newcommand{\NMSU}{New Mexico State University (NMSU), Las Cruces, NM, 88003, USA}
\newcommand{\Otterbein}{Otterbein University, Westerville, OH, 43081, USA}
\newcommand{\Oxford}{University of Oxford, Oxford OX1 3RH, United Kingdom}
\newcommand{\PNNL}{Pacific Northwest National Laboratory (PNNL), Richland, WA, 99352, USA}
\newcommand{\Pitt}{University of Pittsburgh, Pittsburgh, PA, 15260, USA}
\newcommand{\StMarys}{Saint Mary's University of Minnesota, Winona, MN, 55987, USA}
\newcommand{\SLAC}{SLAC National Accelerator Laboratory, Menlo Park, CA, 94025, USA}
\newcommand{\SDSMT}{South Dakota School of Mines and Technology (SDSMT), Rapid City, SD, 57701, USA}
\newcommand{\Syracuse}{Syracuse University, Syracuse, NY, 13244, USA}
\newcommand{\TelAviv}{Tel Aviv University, Tel Aviv, Israel, 69978}
\newcommand{\Tennessee}{University of Tennessee, Knoxville, TN, 37996, USA}
\newcommand{\UTA}{University of Texas, Arlington, TX, 76019, USA}
\newcommand{\Tubitak}{TUBITAK Space Technologies Research Institute, METU Campus, TR-06800, Ankara, Turkey}
\newcommand{\Tufts}{Tufts University, Medford, MA, 02155, USA}
\newcommand{\VTech}{Center for Neutrino Physics, Virginia Tech, Blacksburg, VA, 24061, USA}
\newcommand{\Warwick}{University of Warwick, Coventry CV4 7AL, United Kingdom}
\newcommand{\Yale}{Yale University, New Haven, CT, 06520, USA}

\affiliation{\Bern}
\affiliation{\BNL}
\affiliation{\Cambridge}
\affiliation{\Chicago}
\affiliation{\Cincinnati}
\affiliation{\CSU}
\affiliation{\Columbia}
\affiliation{\Davidson}
\affiliation{\FNAL}
\affiliation{\Harvard}
\affiliation{\IIT}
\affiliation{\KSU}
\affiliation{\Lancaster}
\affiliation{\LANL}
\affiliation{\Manchester}
\affiliation{\MIT}
\affiliation{\Michigan}
\affiliation{\NMSU}
\affiliation{\Otterbein}
\affiliation{\Oxford}
\affiliation{\PNNL}
\affiliation{\Pitt}
\affiliation{\StMarys}
\affiliation{\SLAC}
\affiliation{\SDSMT}
\affiliation{\Syracuse}
\affiliation{\TelAviv}
\affiliation{\Tennessee}
\affiliation{\UTA}
\affiliation{\Tubitak}
\affiliation{\Tufts}
\affiliation{\VTech}
\affiliation{\Warwick}
\affiliation{\Yale}

\author{C.~Adams} \affiliation{\Harvard}
\author{M.~Alrashed} \affiliation{\KSU}
\author{R.~An} \affiliation{\IIT}
\author{J.~Anthony} \affiliation{\Cambridge}
\author{J.~Asaadi} \affiliation{\UTA}
\author{A.~Ashkenazi} \affiliation{\MIT}
\author{M.~Auger} \affiliation{\Bern}
\author{S.~Balasubramanian} \affiliation{\Yale}
\author{B.~Baller} \affiliation{\FNAL}
\author{C.~Barnes} \affiliation{\Michigan}
% ...

\collaboration{The MicroBooNE Collaboration} \email{microboone\_info@fnal.gov} \noaffiliation

	\date{\today}
	\begin{abstract}	
	
		\vspace{25pt}		
		Cosmic ray (CR) interactions can be a challenging source of background for neutrino oscillation and cross-section measurements in surface detectors.
		We present methods for CR rejection in measurements of charged--current quasielastic--like (CCQE--like) neutrino interactions,
		with a muon and a proton in the final state,
		measured using liquid argon time projection chambers (LArTPCs).
		Using a sample of cosmic data collected with the \uB\ detector,
		mixed with simulated neutrino scattering events,
		a set of event selection criteria is developed that produces an event sample with minimal contribution from
		CR background.
		Depending on the selection criteria used a purity between $50\%$ and $80\%$ can be achieved
		with a signal selection efficiency between $50\%$ and $25\%$, with higher purity coming at the expense of lower efficiency.
		While using a specific dataset from the \uB\ detector and selection criteria values optimized for CCQE-like events,		
		the concepts presented here are generic and can be adapted for various studies of exclusive \neumu\ interactions  in LArTPCs.
		
	\end{abstract}
	\maketitle
	
	\clearpage


	\section{Introduction}
		\label{sec:Introduction}
	Liquid argon time projection chambers (LArTPCs) \cite{Rubbia:1977zz} serve as both the target and the detection medium for several operating and planned neutrino oscillation experiments 
	\cite{Antonello:2015lea,Abi:2018dnh}.
	The main advantage of using liquid argon TPCs in neutrino experiments is their fine--grained spatial resolution and precise charge measurements that allow for a low-threshold,
	three-dimensional reconstruction of charged particles and photons produced in neutrino interactions.
	
	The final-state detection capabilities of LArTPCs make them excellent detectors for next generation neutrino oscillation experiments,
	such as searches for sterile neutrinos in short--baseline oscillations \cite{Antonello:2015lea},
	and determination of the neutrino mass ordering and CP violating phase of the neutrino mixing matrix in
	long--baseline oscillation measurement analyses \cite{Abi:2018dnh}.
	Their superb final--state reconstruction capabilities also make them ideal for studies of neutrino--nucleus interaction cross-sections that serve as an important input to many oscillation measurements and searches for new physics.
	In addition,
	liquid argon TPCs have been proposed as possible detectors for proton decay,
	neutrinos from core--collapse supernovae,
	solar neutrinos,
	the diffuse supernova neutrino flux \cite{Abi:2018dnh},
	and dark matter \cite{Aalseth:2017fik}.
	
	The Short Baseline Neutrino (SBN) program at Fermilab
	is composed of a three-detector complex,
	all using LArTPCs \cite{Antonello:2015lea}.
	\uB\ \cite{Acciarri:2016smi} is the first detector commissioned and has been taking data since October 2015.
	Its primary goal is to address observations of an excess of low--energy electron--neutrino like neutrino interactions \cite{Aguilar-Arevalo:2013pmq}
	that could give hints of physics beyond the Standard Model,
	such as sterile neutrino interactions.
	Achieving the goals of SBN and the future DUNE experiment \cite{Abi:2018dnh} requires a precise understanding of neutrino-argon interaction cross sections,
	as well as detailed knowledge of nuclear final--state interactions (FSI) and other nuclear effects.	
	The excellent particle reconstruction capabilities of LArTPCs allow them to detect neutrino interactions using exclusive final states
	that allow for the required understanding of the neutrino--nucleus interaction and impact of nuclear effects.
	
	An example of such a process is charged-current quasielastic (CCQE) scattering \cite{LlewellynSmith:1971uhs} off a bound neutron in a nucleus.
	We choose events in which the neutrino interacts with a single intact nucleon from the nucleus,
	without producing any additional particles.
	This is one of the simplest lepton--nucleus interactions in the energy regime relevant for most neutrino oscillation experiments.
	Measurements of such interactions also allow benchmarking theoretical models of the neutrino interaction cross-section,
	needed for the precision extraction of oscillation parameters.
	Therefore,
	CCQE interactions are an appealing neutrino--interaction channel for use in high-precision oscillation measurements \cite{e4v,Mosel:2013fxa}.
	
	A typical working definition of CCQE interactions in LArTPCs requires the reconstruction of a vertex with exactly one muon, one proton, and no additional particles.
	Such a classification can include contributions from non--CCQE interactions that lead to production of additional particles that are below the detection threshold
	or are absent from the final state due to nuclear effects,
	e.g., pion absorption;
	we refer to these as ``CCQE-like'' events. 
	A significant challenge in studying such interactions stems from cosmic rays whose interactions in
	a LArTPC can mimic CCQE-like interaction topologies.
	This is especially significant for detectors located on the Earth's surface,
	where such cosmic ray (CR) interactions are vastly more abundant than neutrino interactions. 
	
	There are various ways by which cosmogenic background particles can mimic CCQE--like interactions,
	including: 
	a single CR traversing the detector and being misidentified as two 
	trajectories with a common vertex due to local non-active regions in the detector;
	cosmogenic neutron interactions in the detector that result in the production of multiple charged particles in the final state;
	muon decays that result in the production of Michel electrons;
	and CR scattering off an argon nucleus misidentified as two charged particles with a common vertex.
	Cosmic rays can also produce electromagnetic activity that can mimic electron--neutrino interactions and thus impact \neue\ measurements.

	This article presents a collection of cosmogenic background rejection methods for studies of muon--proton pairs from CCQE--like neutrino interactions within a LArTPC.
	We use a mixed sample of cosmic data collected
	with the \uB\ detector and simulated neutrino interactions generated using the GENIE Monte Carlo (MC) event generator \cite{Andreopoulos:2015wxa}. 
		While focusing on muon--proton final states and using CR data collected by the \uB\ detector,
	the methods presented here can be adapted for other studies using LArTPC detectors and different exclusive neutrino interactions,
	such as charged pion production, e.g. $\nu_{\mu} n \to \mu^{-} \Delta^{+} \to \mu^{-} n \pi^{+}$,
	where only the muon and the pion emerge from the nucleus.

	This article is organized as follows.
	In Sec. \ref{sec:CaseStudyMicroBooNE},
	we briefly present the \uB\ detector \cite{Acciarri:2016smi},
	describe its trigger and the effect on the accepted event rates,
	and present the event sample used in this study.
	In Sec. \ref{sec:DetectorObservables},
	we present a collection of event selection criteria,
	denoted as ``cuts'',
	for CCQE--like studies that are based on the detector response,
	and examine their impact on CR data and the simulated neutrino signal.
	In Sec. \ref{sec:KinematicalSignature},
	we present complementary kinematic cuts based on known properties of CCQE interactions.
	Finally,
	in Sec. \ref{sec:Cosmic rejection with the above criteria},
	we present the sequential impact of each cut on the CR background rejection and neutrino--interaction selection signal efficiency and purity.

	We note that this work focuses on CR background rejection using real data.
	The impact of the event selection criteria on the simulated neutrino signal is shown as a reference.

	\section{Experimental Setup}
	\label{sec:CaseStudyMicroBooNE}
	
\begin{figure*}[htb]
    		\includegraphics[trim={0cm 0cm 0cm 0cm},clip,width=0.4\linewidth]{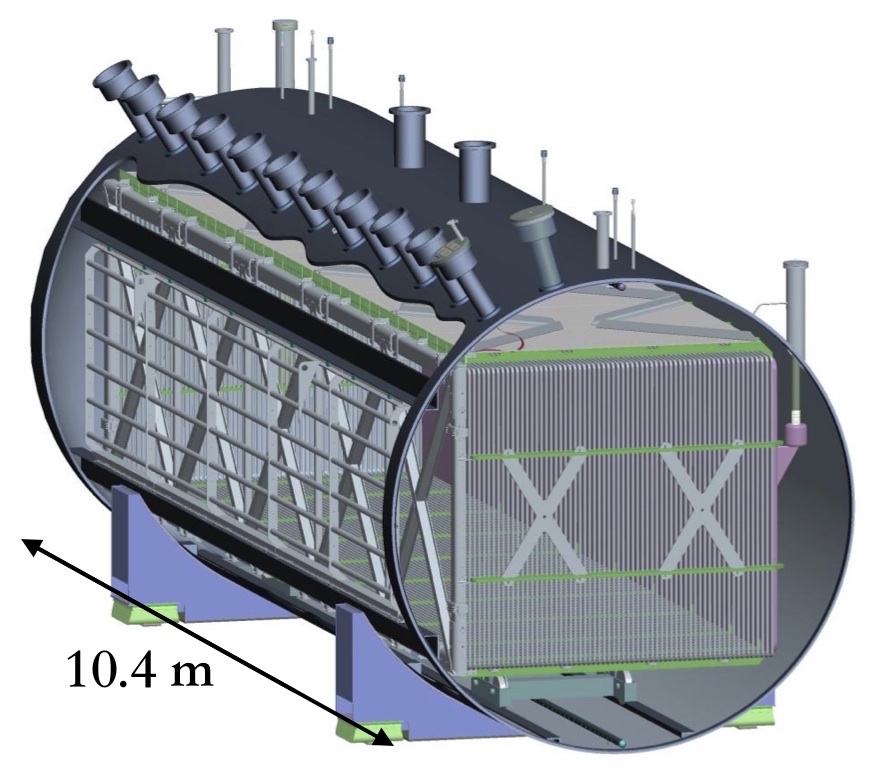} 
    		\includegraphics[width=0.5\linewidth]{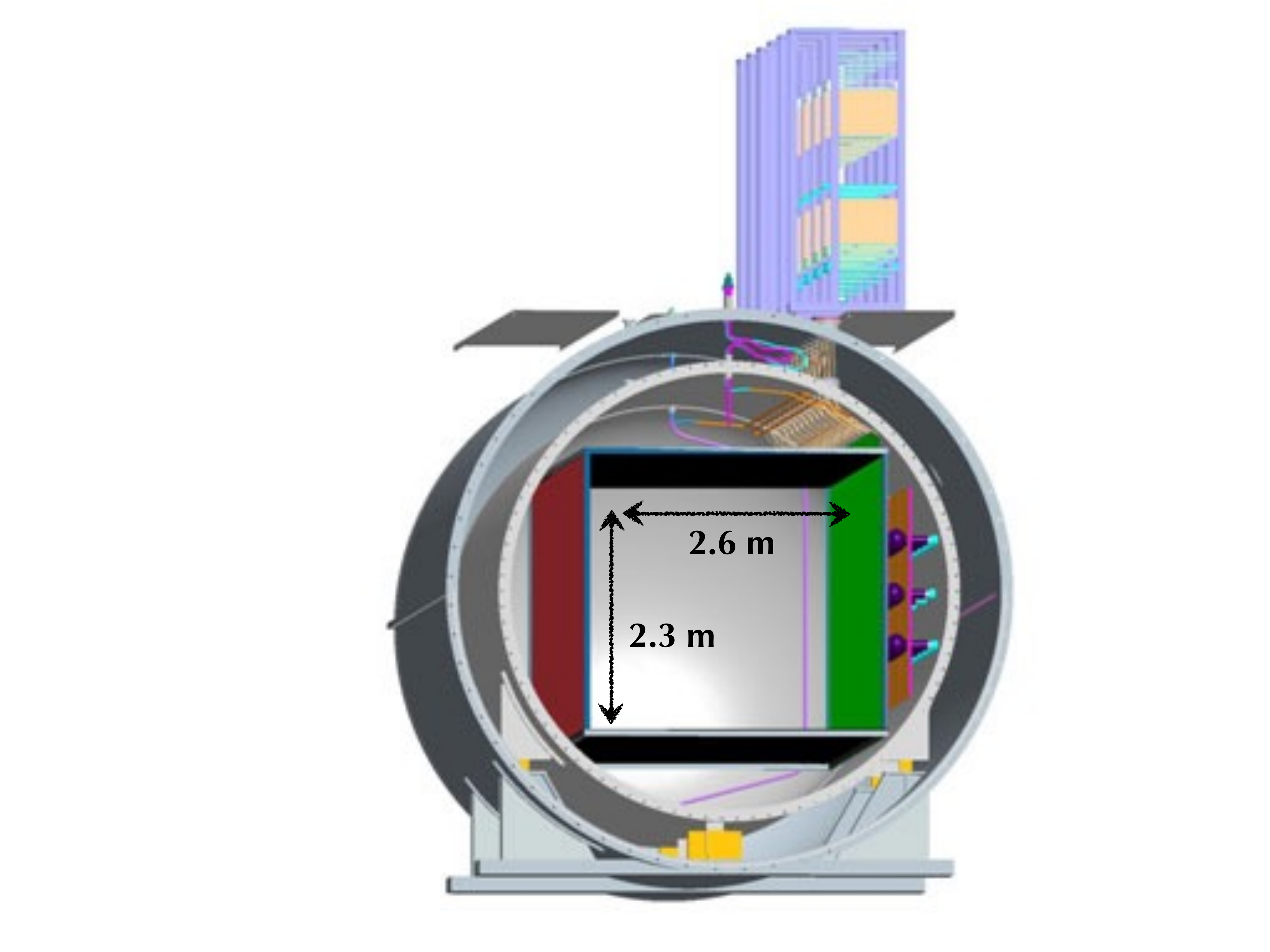} 		
		\caption{A schematic view of the \uB\ detector. 
				The three wire planes,
				shown on the right side of the cylindrical cryostat in the right--hand image,
				contain vertical wires and wires $\pm 60 ^{\circ}$ to the vertical.
				The wire planes are the anode.}
		\label{fig:uBSchematic}
 	\end{figure*}

			The \uB\ detector is a 170 tonne LArTPC with an active mass of 85 tonnes.
 	It is located 463 meters downstream from the production target of the Booster Neutrino Beam (BNB) at FNAL.
	The BNB energy spectrum ranges up to 2 GeV and peaks around 0.7 GeV.
	See \cite{Acciarri:2016smi} for details.
	The detector consists of a rectangular cuboid TPC with dimensions of 256~cm (width) $\times$ 232~cm (height) $\times$ 1036~cm (length),
	displayed in Fig. \ref{fig:uBSchematic}.
	The TPC includes three wire readout planes with 3 mm spacing: a vertical collection plane, labeled as Y, and two induction planes,
	labeled as U and V with wires oriented $\pm 60^\circ$ with respect to the vertical.
	We define a right--handed coordinate system in which $\hat{z}$ is along the beam direction, the drift direction $\hat{x}$ is horizontal, and the $\hat{y}$ direction is vertical.
	The origin $(x=0,y=0,z=0)$ is chosen at the the anode plane, the detector vertical center, and the upstream side of the detector,
	respectively.
	The stainless steel cathode
	operates at -70 kV.

	Neutrino interactions occur in the LArTPC when a neutrino from the beam interacts with an argon nucleus.
	The secondary charged particles produced in the interactions travel through the liquid argon,
	lose energy through ionization processes, and leave a trail of ionization electrons.
	These ionization electrons drift to the wire planes in an electric field of 273 V/cm.
	It takes about 2.3 ms for an ionization electron to travel the maximal drift distance from cathode to anode of 256 cm.
	Consequently, 
	the triggered time window of the TPC is opened for slightly more than 4.8 ms.
	The wire planes are biased in such a way that drift electrons merely induce a pulse in the first two (U and V induction) planes and are collected in the plane
	farthest from the cathode,
	the collection plane.
	These signals are used to create three distinct two-dimensional views
	in terms of wire position and time \cite{Adams:2018gbi}.

	Thirty-two photomultipliers tubes (PMTs) are placed outside the TPC volume facing the anode plane to collect the scintillation light
	from the de-excitation of argon dimers produced along with the ionization electrons.
	A TPB plate in front of each PMT absorbs the UV argon scintillation light and re-emits in a wavelength range to which the PMTs are sensitive.
	The PMT signals in coincidence with the BNB beam gate (or an external, off-beam window) is used as a trigger condition and to define the reference time for the event. 
	An ``event'' in \uB\ consists of a continuous readout of the TPC and the associated PMT light during a 4.8 ms drift window.

	During data-taking,
	a software trigger rejects events with light levels below that expected for a neutrino interaction.
	To enrich the recorded event sample with neutrino induced events in the beam window, 
	we implement an additional trigger in the data-acquisition software during event building which scans the PMT waveforms within the beam and external windows.
	If the sum of all PMT waveforms exceeds a pre-set threshold equivalent to $6.5$ photo-electrons (PEs),
	in any 100 ns time window within the 1.6 $\mu$sec beam gate,
	the event is saved.
	To open this time window and sum the PMTs signals,
	a 0.5 PE signal is required to be recorded in at least one individual PMT.
	
	At nominal beam intensity and without a PMT trigger,
	one in roughly 500 BNB beam spills is expected to contain a neutrino interaction.
	The PMT trigger provides a richer sample in which approximately 1 in 10 spills will contain a neutrino interaction. 
	Even with this PMT trigger,
	further suppression of the CR background by one to two orders of magnitude is required for isolating neutrino interactions.

		The cosmogenic muon rate in \uB\ is estimated to be 5.5 kHz \cite{Acciarri:2017rnj},
	which corresponds to about 20 muons per TPC drift time window of 4.8 ms and constitutes the main source of background to neutrino interaction events.

	To study the cosmogenic background, additional drift windows are recorded during periods between neutrino beam spills,
	or when the Fermilab accelerator is off.
	This acquisition window,
	being out of time with the neutrino beam,
	is referred to as the external trigger and is configured to be superseded by beam triggers in the event of overlapping acquisition windows.

	\subsection{The event sample used for this study}
	
	For CR background studies we use externally triggered data, during which the beam is off, as mentioned above.
	Each real CR data event is mixed with a simulated neutrino interaction event produced by GENIE \cite{Andreopoulos:2015wxa}.
	The overlay is performed such that each simulation event is combined with a single external triggered event.
	Propagation of final state particles following the simulated neutrino interactions is simulated using GEANT4 \cite{geant}.
	Signals along the wires from ionization electrons are simulated using the LArSoft simulation \cite{Snider:2017wjd}.
	
	The sample contains about $10^6$ cosmic data events and the same number of simulated events,
	which is equivalent to that expected for about $10^{21}$ protons on target (POT) of data.
	In \uB\ triggered data,
	fewer than $10\%$ of the triggered events contain a neutrino interaction;
	the true mixture of CR-neutrino data should be one simulated signal for about every ten beam triggered events.
	The external triggered events do not have the PMT trigger applied,
	 and thus represent an unbiased sample of cosmic data.
			
	We define the CCQE--like signal as events with a muon and a proton in the final state, both fully contained within the TPC, 
	that originate from charged-current \neumu-Ar scattering.
	We look for events with exactly one proton with momentum greater than 200 MeV/c and any number of protons below 200 MeV/c;
	We denote these events as \CCIpOpi.
	Our signal definition allows any number of neutrons at any momenta,
	any number of charged pions with momentum lower than 70 MeV/c (about 3 cm track length in liquid argon),
	and no neutral pions,
	electrons or photons (at any momentum).
	The minimal proton momentum requirement is due to its stopping range in LAr,
	which is about 6 mm,
	corresponding to two 3 mm wire pitches of the TPC.
		
	\subsection{Track reconstruction and cosmic background rejection prior to vertex building}
	\label{sec:Track reconstruction in LArSoft}
	\label{sec:Cosmic rejection prior to this analysis}
	
	Track reconstruction consists of three main stages:
	hit reconstruction,
	candidate CR track reconstruction (PandoraCosmic),
	and candidate neutrino--induced track reconstruction (PandoraNu) \cite{Acciarri:2017hat}.
	The first stage includes reconstructing individual hits on the TPC wires.
	In the second stage,
	the PandoraCosmic algorithm
	attempts to combine hits to construct as many candidate CR tracks as possible,
	identified by their geometric information, e.g. downward tracks.
	The hits associated with identified CR tracks are removed from further analysis.
	In the third step,
	the PandoraNu algorithm is used for track reconstruction from the remaining set of hits \cite{Acciarri:2017hat}.
	
	The impact of the PandoraCosmic removal pass was studied by comparing two simulated samples of cosmic events,
	with and without the application of the PandoraCosmic removal pass.
	We found it rejects about $20\%$
	of track pairs,
	induced by CRs,
	that might imitate neutrino interaction events.
		

	\section{Event selection based on detector observables}
	\label{sec:DetectorObservables}
	
	The LArTPC response to CCQE--like neutrino interaction is different from its response to pairs of reconstructed tracks induced from CRs,
	as the former produces a muon and a proton with typical momenta of a few hundred MeV/c primarily traveling horizontally,
	while the latter consists primarily of vertically traveling high-momentum muons.
	Consequently,
	we can reject CR backgrounds using information related to the energy of the tracks produced in CCQE--like neutrino interactions
	(i.e., \CCIpOpi\ events),
	the distance between the tracks and the scintillation light produced,
	the length of the muon and proton tracks,
	and the location of vertices within a fiducial volume.
	This section describes these cuts and quantifies their impact on the cosmic data rejection and simulated neutrino signal loss.
	
	The set of cuts presented in this section rely on low-level detection quantities, 
	which we find are well-modeled by the \uB\ detector simulation \cite{Adams:2018gbi}.
	As a result,
	these methods represent a robust prescription for isolating \mup\ pairs primarily originating from neutrino--argon interactions.

	\setlength{\belowdisplayskip}{3pt} \setlength{\belowdisplayshortskip}{3pt}
	\setlength{\abovedisplayskip}{3pt} \setlength{\abovedisplayshortskip}{3pt}

	\subsection{Close track identification and reconstruction efficiencies}
	
	We consider pairs of tracks that are fully contained in the fiducial volume of \uB,
	defined by
	\begin{equation}
	\label{eq:FV}
	\begin{array}{rllll}
		3 & < & x & < &253 \textrm{ cm},  \; \\
		-110 & < & y & < &110 \textrm{ cm}, \;  \\
		 5 &< & z &< & 1031 \textrm{ cm}. 
	\end{array}
	 \end{equation}		 
	 Tracks are fully contained if both beginning and endpoints are within this volume.
	 
	\begin{figure}[h]
  		\centering
    		\includegraphics[width=\linewidth]{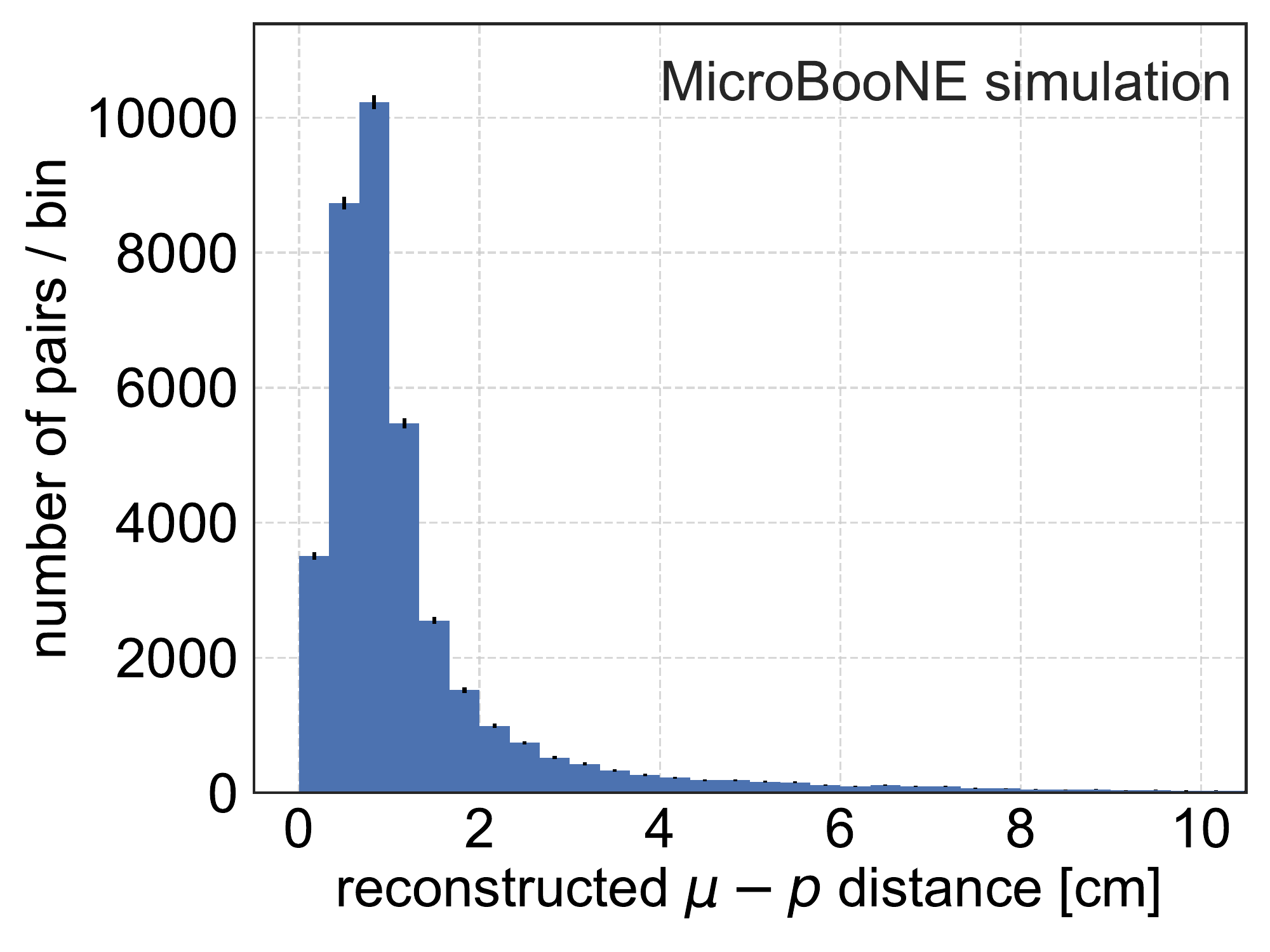} 
		\caption{The reconstructed distance between the start point of the muon and the proton tracks in simulated \CCIpOpi\ events in \uB.
		The error bars are statistical only.}
		\label{fig:mup_distance}
 	\end{figure}

	To determine the desired proximity between the start points of the two tracks,
	we study the distribution of the reconstructed three--dimensional distance between the start points of the $\mu$ and $p$ tracks in all simulated \CCIpOpi\ events.
	The resulting distribution is shown in Fig. \ref{fig:mup_distance}.
	The resolution for the start point of a track is on the order of 1 cm.
	We conservatively include vertices in which the muon and proton candidate track start points are closer than 11 cm,
	to minimize possible detector model dependence.

	The fraction of \CCIpOpi\ events with a muon--proton vertex in the TPC active volume, contained tracks,
	and no additional track detected within 11 cm of the reconstructed vertex,
	is about $20\%$.
	We denote their selection as the ``preselection'' stage.
	However, 
	this preselection alone does not remove any ``broken" CR trajectories,
	where a single particle from a CR is reconstructed as two tracks,
	typically characterized by a separation of the start points less than 11 cm.

	The reconstruction efficiency for cosmic--induced track--pairs,
	as well as the efficiency in reconstructing an artificial pair of tracks at close proximity,
	were studied using a set of simulated cosmic events generated using the CORSIKA generator \cite{Heck:1998vt}.
	On average,
	for each event,
	there are $0.32 \pm 0.02$ pairs of contained PandoraNu CR tracks reconstructed with a common start point (closer than 11 cm).
	About $75 \%$ of these pairs originate from mis-reconstructed broken tracks,
	$5 \%$ from misconstruction of the trajectories of two distinct particles with intersecting trajectories,
	and $20 \%$ from mis--interpreted muon--electron pairs due to Michel decays of CR muons.
	These ``fake'' vertices are reduced by the cuts discussed below.
	
	\subsection{Energy deposition profile of the proton and muon candidates}
	\label{sec:Energy deposition profile of the proton and muon candidates}
	
	Particle identification (PID) can be used to discriminate broken muon tracks from real muon-proton pairs. PID is based on energy deposition.
	Particle identification in LArTPCs is usually done using calorimetry as measured by the energy deposition profile $dE/dx$ along the track,
	as a function of its residual range (the distance of each trajectory point from the end of the track).
	
	\begin{figure}[h]
  		\centering		
			\begin{tabular}{@{}p{0.48\linewidth}@{\quad}p{0.48\linewidth}@{}}
		 	\subfigimg[width=\linewidth]{(a)}{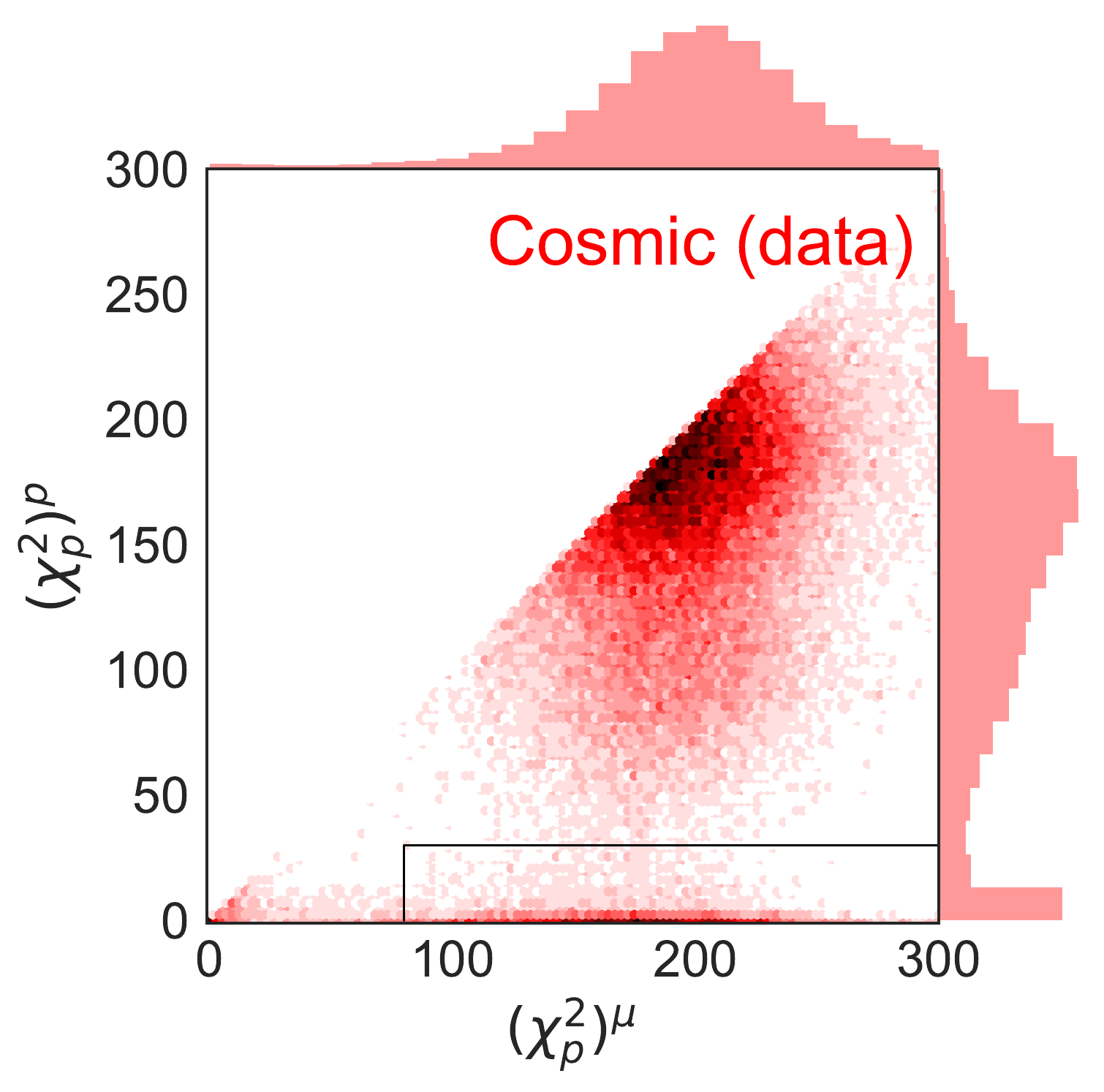}{2.5}{30} &
		 	\subfigimg[width=\linewidth]{(b)}{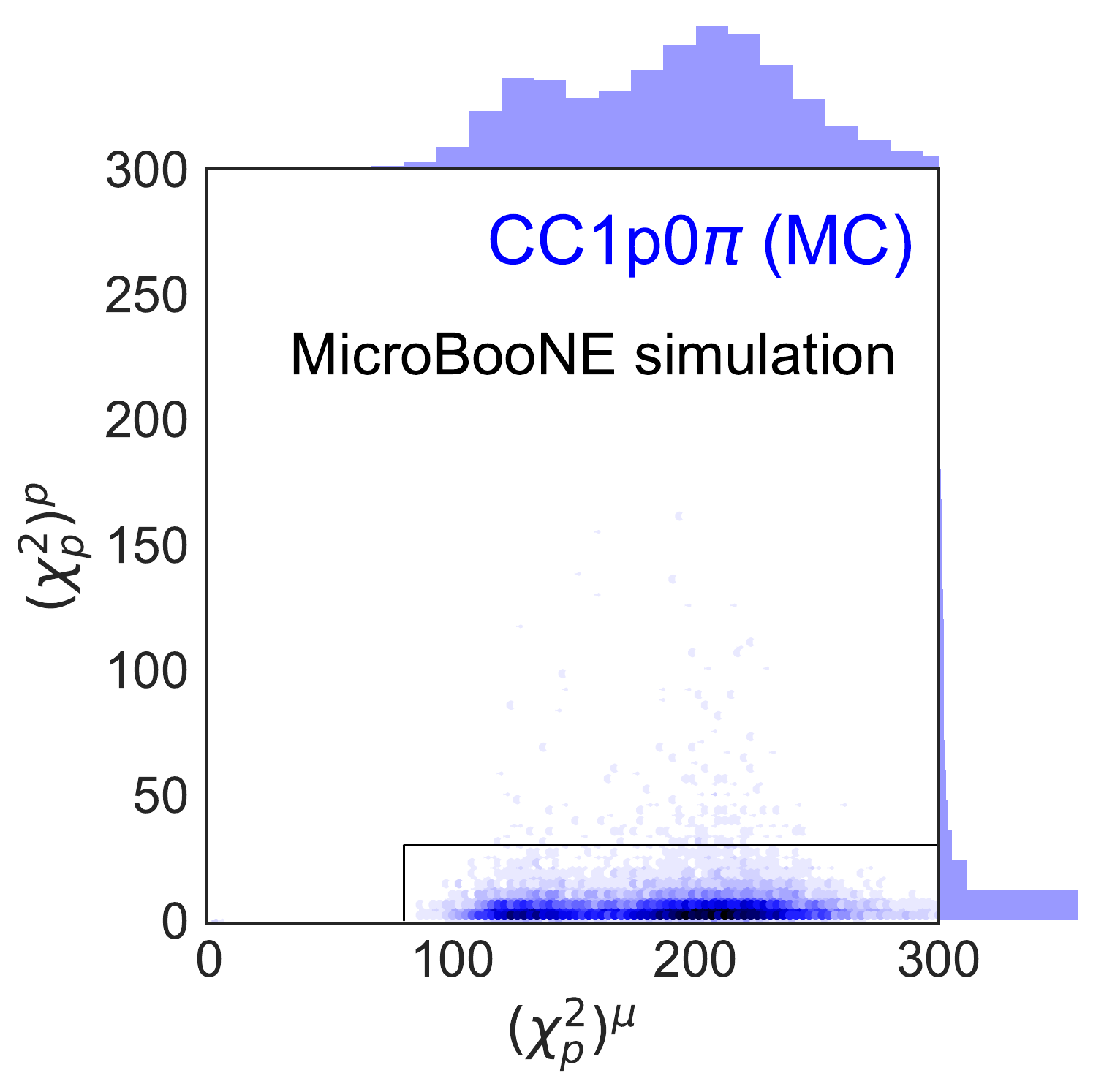}{2.5}{30} \\
		 	\subfigimg[width=\linewidth]{(c)}{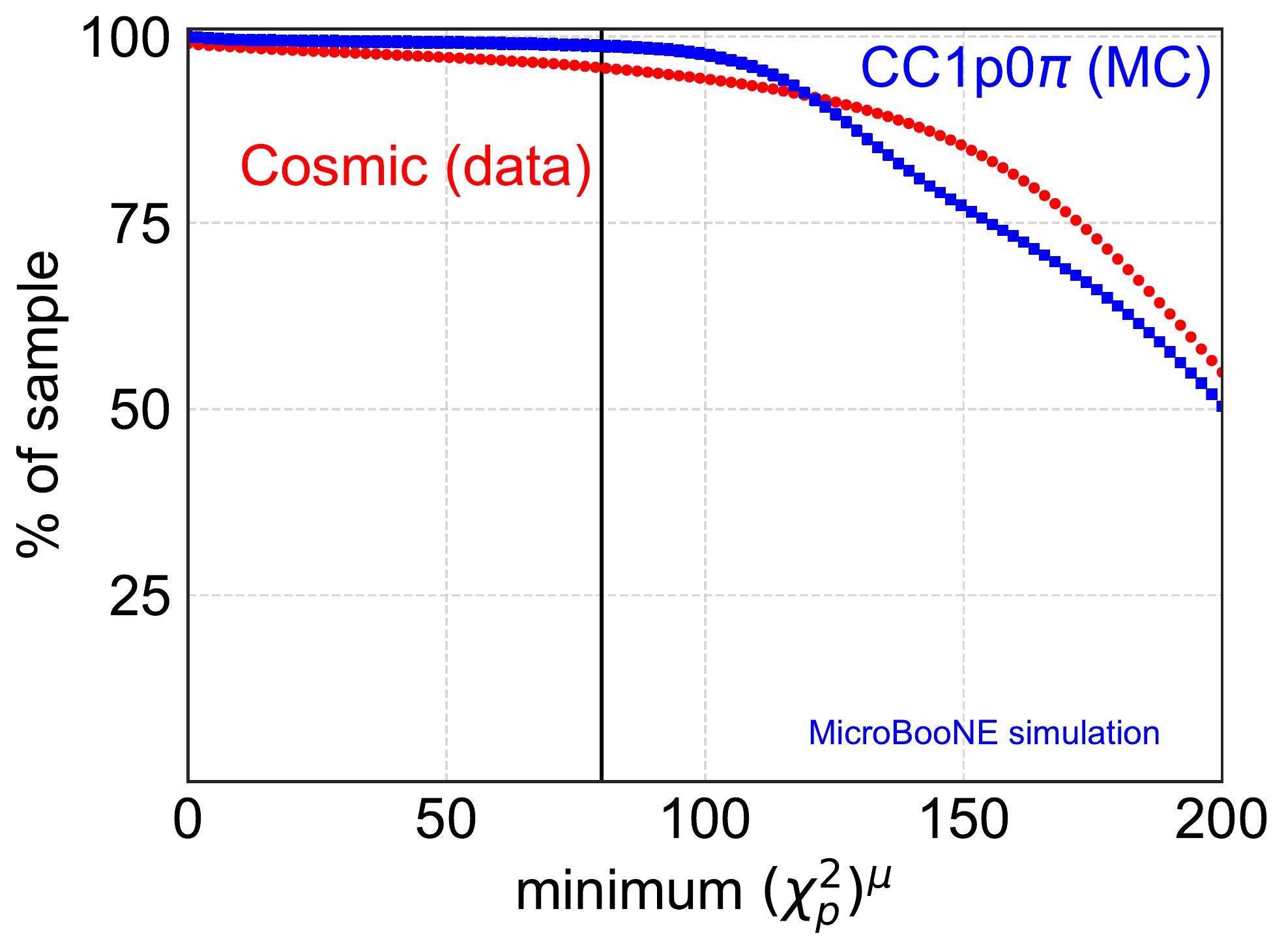}{3}{30} &
		 	\subfigimg[width=\linewidth]{(d)}{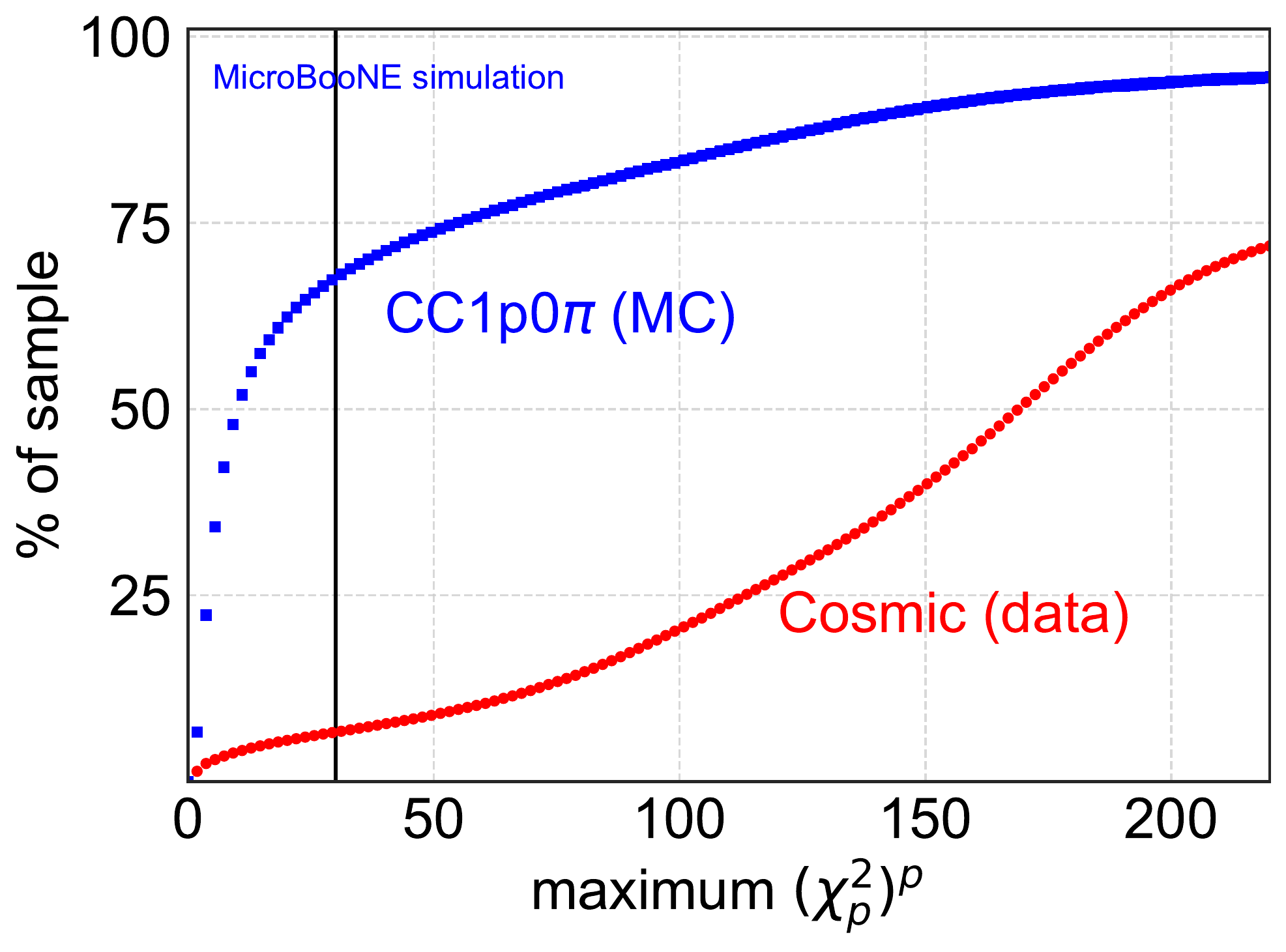}{3.2}{103} \\
			\end{tabular}
		\caption{The distribution of \ChiSqrP\ of the proton vs. the muon candidate for two identified tracks in a close-proximity pair in \uB,
			for (a) cosmic data and 
			(b) \CCIpOpi\ simulated signal.
			The proton candidate is labeled as the track with the smaller \ChiSqrP. 
			Also shown are the one-dimensional projections of the distributions.
			A cut of \CutChiSqrP\ to suppress CR background contribution is depicted in the figure.
			(c) The impact of a cut on $(\chi^{2}_{\textrm{p}})^{\mu}$ on the number of simulated signal \CCIpOpi\ events lost and the number of CR background events rejected. 
			(d) The impact of a cut on $(\chi^{2}_{\textrm{p}})^{p}$
				on the number of simulated \CCIpOpi\ events lost and CR background events rejected. 
			}
	\label{fig:Chi2Proton_mu_p}
 	\end{figure}
	
	PID is implemented here by comparing the measured \dEdx\ profile to the Bethe-Bloch expectation for a proton (or a muon) in liquid argon using a $\chi^2$ test,
	where we normalize  $\chi^2$ to  the number of hits in the track.
	We assume two tracks reconstructed at close proximity are a muon--proton pair
	and label the proton candidate as the tracks with the smaller \ChiSqrP\ (i.e. $\chi^2$  compared to the proton expectation).
	Figure \ref{fig:Chi2Proton_mu_p} shows the distributions of the \ChiSqrP\ of the proton candidate track ($(\chi^{2}_{p})^{p}$)
	vs. \ChiSqrP\ of the muon candidate track ($(\chi^{2}_{p})^{\mu}$) for the mixed cosmic data and simulated signal sample.
	
	As can be seen,
	muon--proton pairs coming from neutrino interactions populate a region where \ChiSqrP\ is very low ($<30$),
	whereas CR pairs have a \ChiSqrP\ around 200.
	This is due to the fact that a non-negligible fraction of
	the latter consist of electromagnetic debris,
	such as delta rays and shower fragments,
	which produce very little ionization in liquid argon. 
	In addition,
	there is a population of events for which the calorimetric reconstruction has failed for at least one of the particles.
	We remove these events with a quality cut requiring $(\chi^2_p)^p>0$.

	Requiring \CutChiSqrP\ suppresses the CR background by about a factor of $20$,
	at the cost of losing about $34\%$ of the \CCIpOpi\ events.

	\begin{figure}[h]
  		\centering
		 \begin{tabular}{@{}p{0.48\linewidth}@{\quad}p{0.48\linewidth}@{}}
		 \subfigimg[width=\linewidth]{(a)}{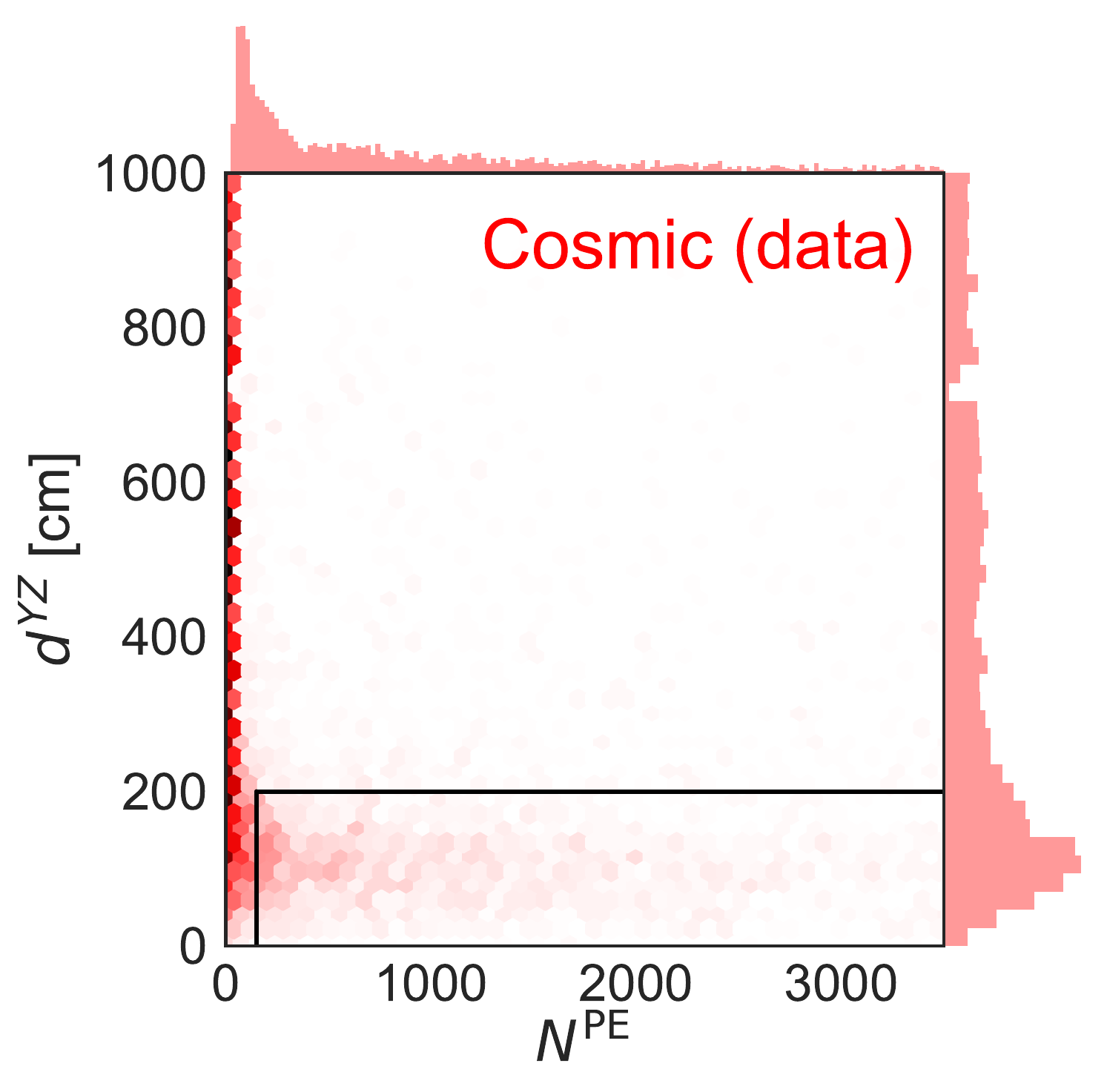}{2.5}{30} &
		 \subfigimg[width=\linewidth]{(b)}{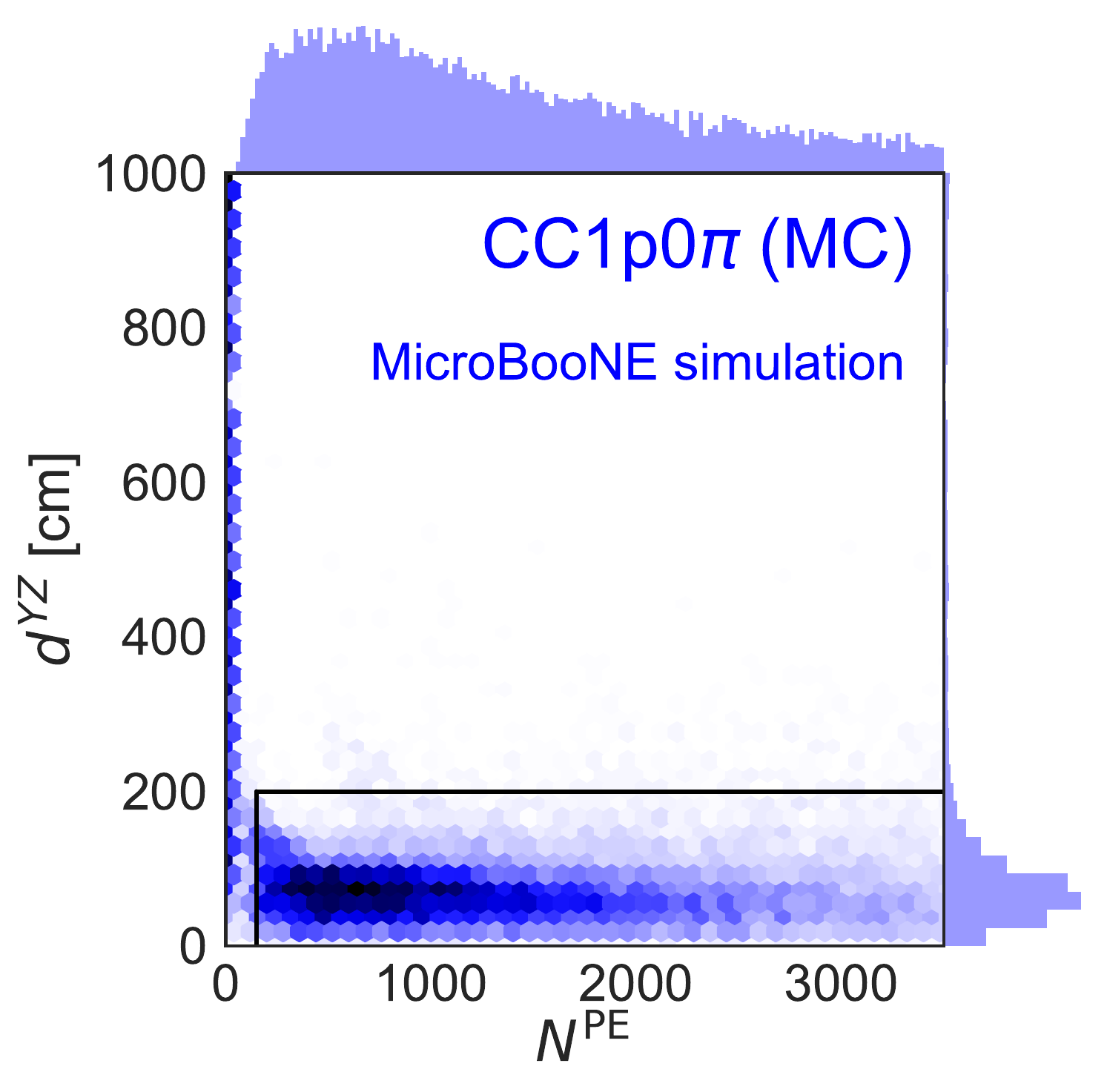}{2.5}{30} \\
		\end{tabular}
		\caption{The distribution of the two-dimensional distance of the reconstructed vertex to the associated flash,
				vs. the number of photoelectrons (PE) recorded in the flash in:
				(a) cosmic data, and (b) simulated \CCIpOpi\ events.
				Also shown are the one-dimensional projections of the distributions.
				}
	\label{fig:MatchedFlash_after_dEdx_cut}
 	\end{figure}

		\begin{figure}[h]
  			\centering
			\begin{tabular}{@{}p{0.48\linewidth}@{\quad}p{0.48\linewidth}@{}}
		 	\subfigimg[width=\linewidth]{(a)}{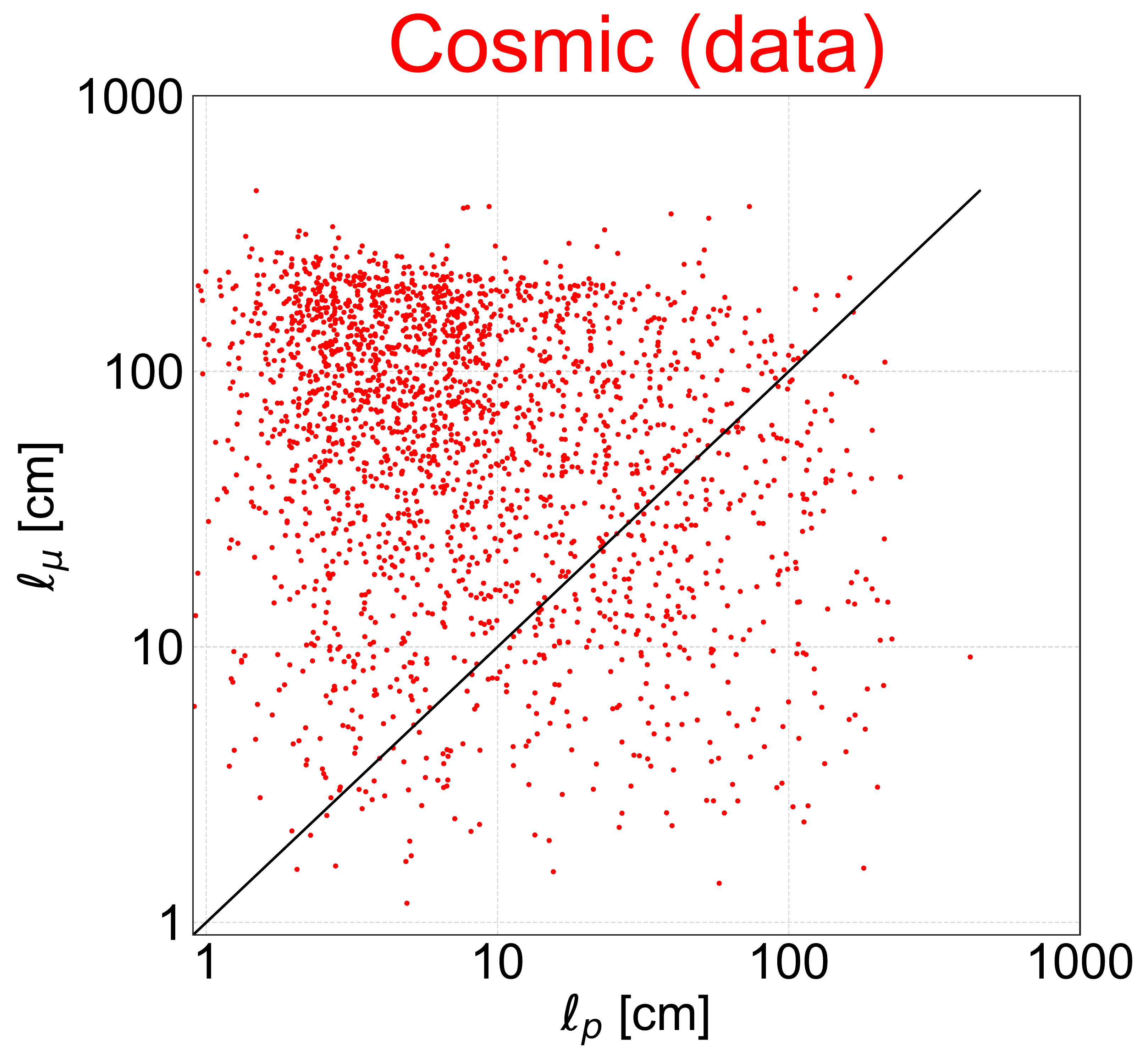}{1.8}{30} &
		 	\subfigimg[width=\linewidth]{(b)}{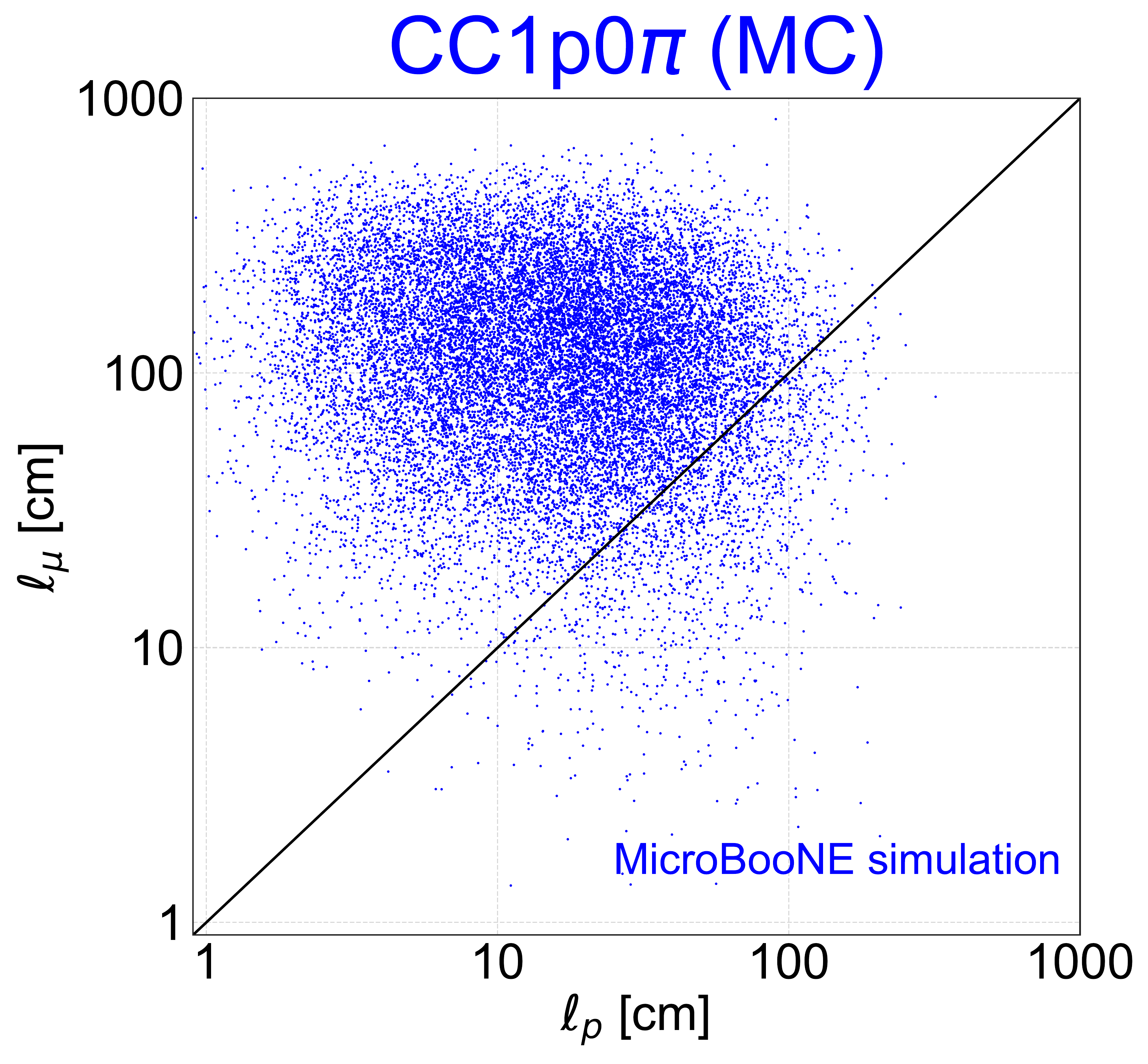}{1.8}{30} \\
			\end{tabular}
			\caption{The distributions of the segmented length of the track labeled as the muon candidate, $l_{\mu}$,
					vs. the track labeled as the proton candidate, $l_{p}$,
					after the application of the cuts on \dEdx\ and matched PMT flash in \uB\ in:
					(a) cosmic data, and (b) simulated \CCIpOpi\ events.} 
			\label{fig:l_mu_p_after_dEdx_and_PMT_cuts}
 		\end{figure}

	\subsection{Optical filtering}
	
	The data from the 32 PMTs spread in the $y-z$ plane behind the TPC wire planes is also useful for CR rejection,
	as CRs typically produce less scintillation light,
	with a larger spread along the tracks,
	as compared to the more localized light produced in the vicinity of the neutrino interactions vertex.
	\uB\ event reconstruction produces a ``flash'',
	if:
	(1) at least three separate PMT pulses exceed the single PE (SPE) level within 30 ns,
	and
	(2) the sum of the three PMT pulses above SPE exceed five PEs.
	The flash includes the sum of all the PMT pulses integrated during an eight microseconds time window from the time of the first coincidence.
	Each event can include several such flashes.
	
	We identify the flash associated with the vertex by comparing the expected PE yield in each PMT to the observed one.
	We use two selection criteria based on this flash:
	(1) the two-dimensional distance of the reconstructed vertex from the center of the associated flash in the $y-z$ plane, $d_{\textrm{YZ}}$,
	and
	(2) the total number of PEs recorded in this flash, $N_{\textrm{PE}}$.
	
	Figure \ref{fig:MatchedFlash_after_dEdx_cut} shows the distribution of $d_{\textrm{YZ}}$ vs $N_{PE}$ for the mixed cosmic data
	and simulated \CCIpOpi\ signal sample.
	The events shown
	have an identified vertex and pass the energy deposition profile cut discussed above.
	Also shown is a cut on \CutFlashYZ\ and \CutFlashPE.
	The results of these cuts is a rejection of the cosmic contribution by about a factor of 2,
	with a signal loss of about $15\%$.

		\subsection{Muon and proton track lengths}
		
	The reconstructed track length is used to further distinguish neutrino interactions from cosmic background.
	Pandora reconstructs segmented lengths of particle tracks, i.e.,
	the accumulated length of multiple straight track segments,
	which do not necessarily sit on a straight line due to multiple coulomb scattering and other interactions.
	Figure \ref{fig:l_mu_p_after_dEdx_and_PMT_cuts} shows the distributions of the segmented length of the muon candidate, $l_{\mu}$,
	vs. the proton candidate, $l_{p}$,
	after application of the cuts on the \dEdx\ profile and optical filtering described above. 
	We expect that the muon tracks will be longer than the proton track with no cuts applied;
	this is indeed predicted to be the case in about $78\%$ of the simulated \CCIpOpi\ events when we identify the proton as the track with smaller \ChiSqrP.
	With the above requirements on the \dEdx\ profile and optical filtering applied,
	this is the case for about $91\%$ of the simulated \CCIpOpi\ events.
	Consequently,
	we require \Cutlmup,
	which results in a signal loss of about $9\%$ and leads to a correct identification of the muon and the proton in about $99\%$ of the remaining pairs.
	The application of this requirement rejects about $20\%$ of the CR background remaining after the previous selection criteria are applied.
				
		\subsection{Broken track removal}
\begin{figure}[h]
		\centering  		
			\begin{tabular}{@{}p{\linewidth}@{\quad}p{\linewidth}@{}}
		 	\subfigimg[width=\linewidth]{(a)}{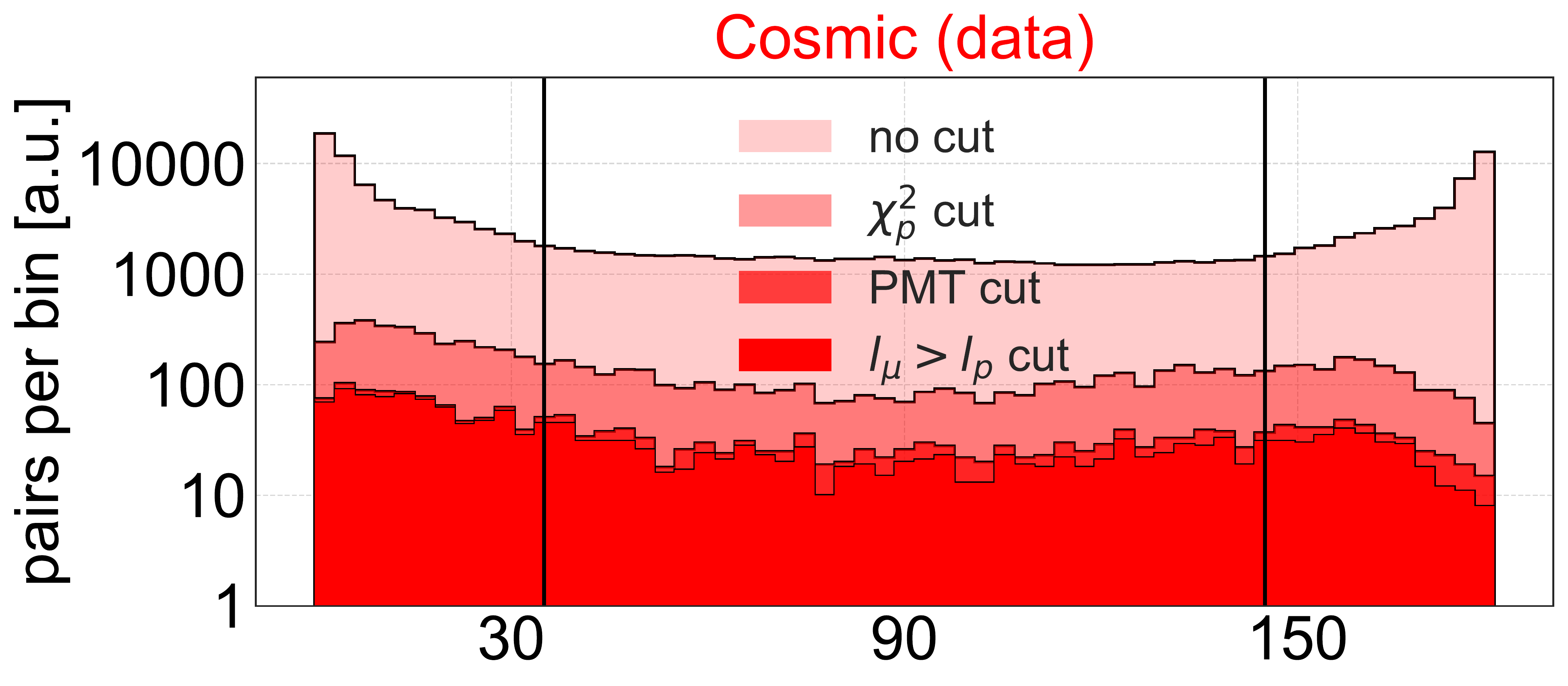}{2}{220} \\
		 	\subfigimg[width=\linewidth]{(b)}{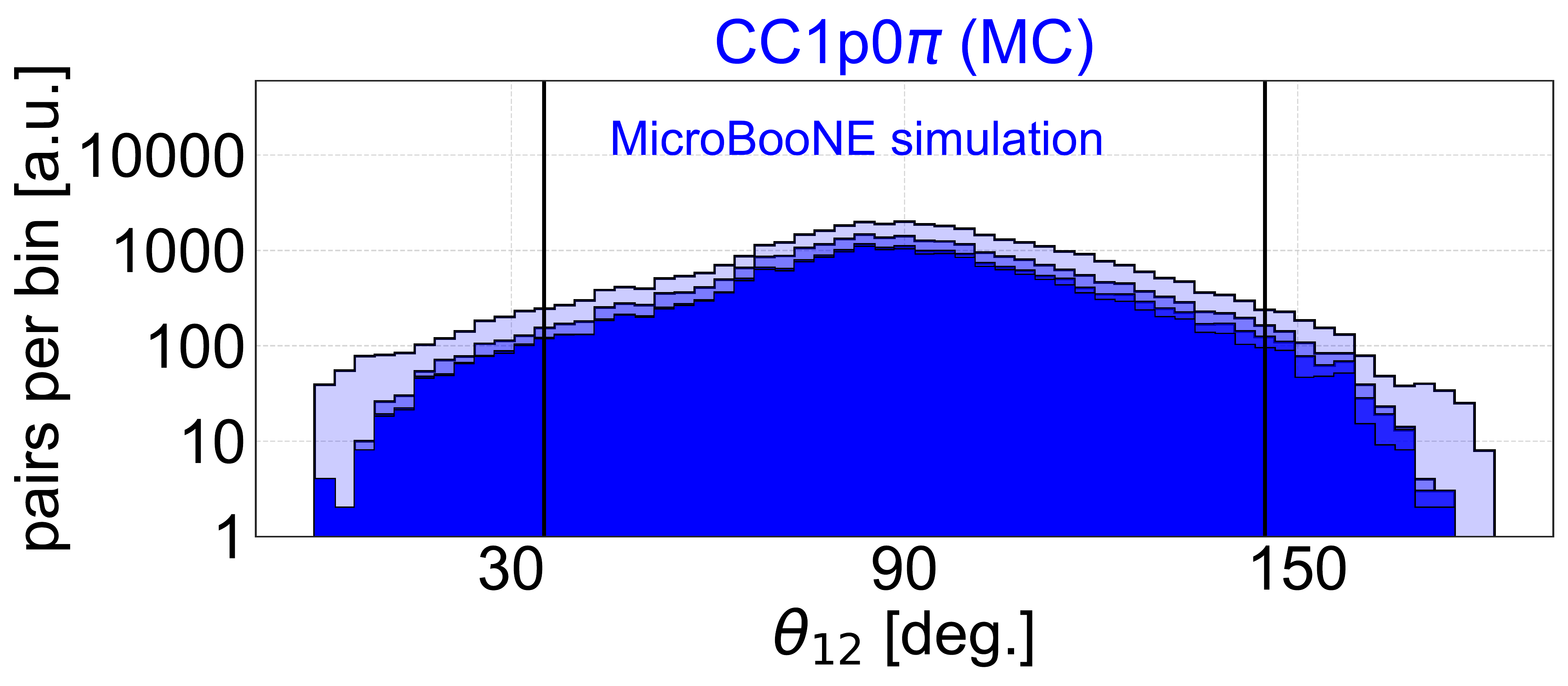}{2}{220} \\
		 	\subfigimg[width=\linewidth]{(c)}{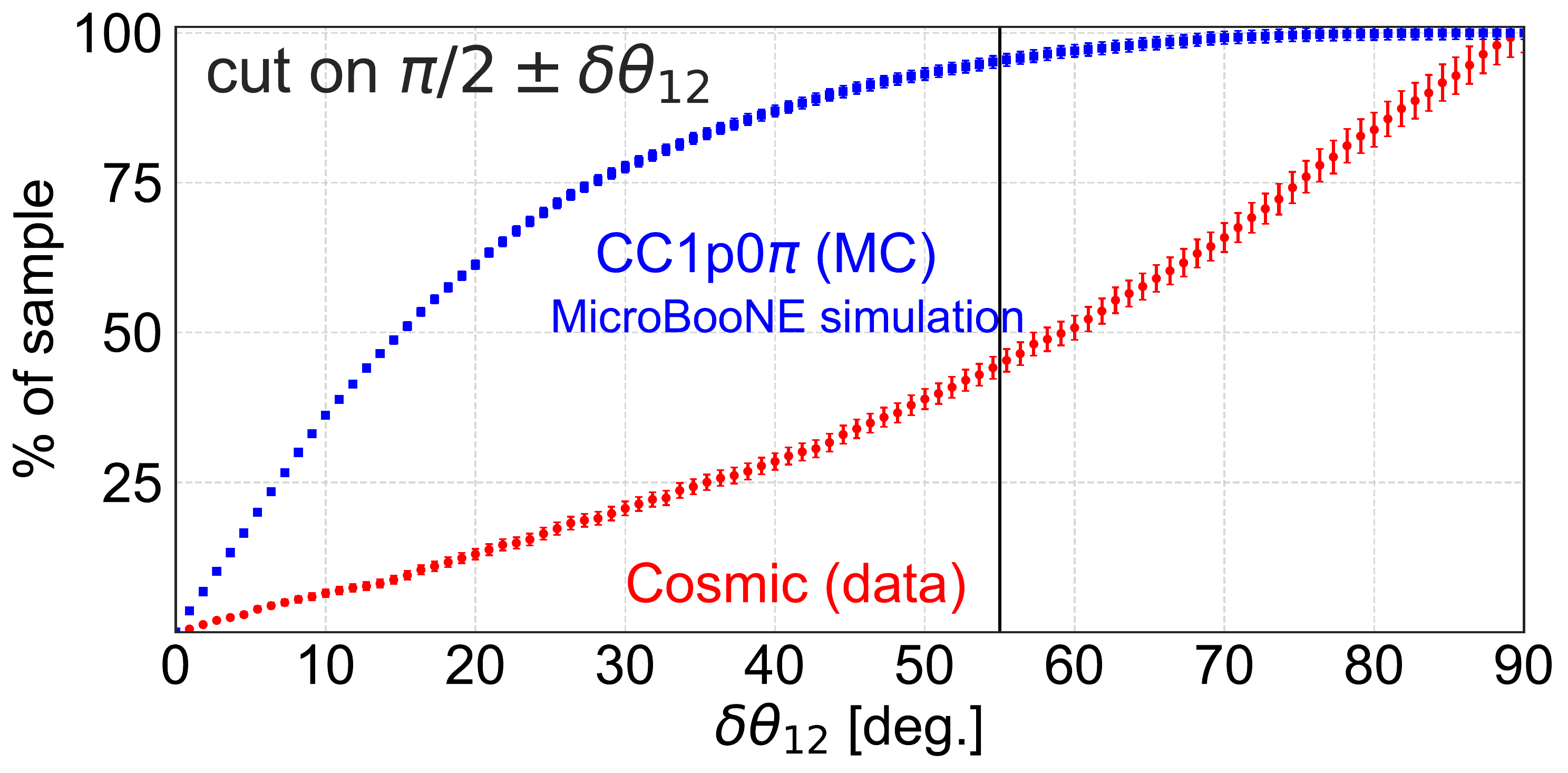}{4.4}{220} \\
			\end{tabular}
		\caption{The distribution of $\theta_{12}$,
				in \uB, for
				(a) CR background data, and (b) \CCIpOpi\ simulated signal.
				A cut of $\delta\theta_{12}=$\CutCollinearity,
				to suppress background contributions is depicted in the figure.
				(c) The impact of a cut on $\delta\theta_{12}$ on the number of simulated signal \CCIpOpi\ events lost and the number of CR background events rejected.}
		\label{fig:theta_12}
		\end{figure}

		As mentioned in Sec. \ref{sec:Introduction},
		some of the 
		two--track events induced by cosmic rays originate from a broken track.
		This may be caused by particles traveling across a region with inactive wires in the TPC,
		a soft scattering of CR muons off the argon nuclei,
		or an inefficiency in hit reconstruction.
		The mitigation of these effects is possible using the collinearity of the two tracks.
		Events originating from broken--tracks can be characterized by having a three-dimensional angle $\theta_{12}$ between the two tracks close to $0^{\circ}$ or $180^{\circ}$.
		This angle is useful in identifying background caused by such tracks.
		Figure \ref{fig:theta_12} shows the individual distributions of the different samples before and after the application of the \dEdx,
		flash,
		and track length selections.
		We require that \CutCollinearity,
		as depicted in the figure.

	\section{Event selection based on quasielastic kinematics}
	\label{sec:KinematicalSignature}

    	The Selection cuts described above are based on the LArTPC detector response and do not rely heavily on the specific two--body kinematical signature of CCQE interactions.
	They lead to a CR suppression of approximately $99.5 \%$
	and maintain a signal purity of about $50\%$.
	Using the two--body kinematics of CCQE interactions,
	such as vertex activity, coplanarity, and the imbalance of the transverse momentum, $p_{T}$,
	allows further CR rejection and increase of the \CCIpOpi\ purity. 
	These variables,
	unlike the detector observables,
	are not modeled by the detector simulation,
	but rely primarily on the model--dependent neutrino interaction generator. 
	This dependence is reduced by using relatively loose cuts and performing cut-sensitivity studies.
	On the other hand,
	the effect of the cuts on the CR background rejection is directly measured with data, and hence contains no such model dependence.
	
	We introduce three relevant cuts that are based on CCQE two--body kinematics.
	The first removes events with a large energy deposit near the vertex,
	where the energy is not associated with the muon and proton tracks. 
	The second is based on the expected coplanarity angle between the plane spanned by the neutrino and muon and that spanned by the neutrino and the proton.
	The third is based on the imbalance of transverse momentum of the final state particles relative to the incoming neutrino direction.
	These cuts can be applied independently or together for enhanced CR rejection by focusing on a specific part of the CCQE--like interaction phase--space.
	
	\subsection{Removing events with large additional energy deposits near the vertex}
		
	Events in which multiple particles are produced but only one muon and one proton track are reconstructed have large charge depositions in the vertex region,
	not associated with the reconstructed muon and proton tracks.
	Such events can be identified and removed to enhance the signal purity.		
	
		\begin{figure*}
    			\includegraphics[width=\linewidth]{{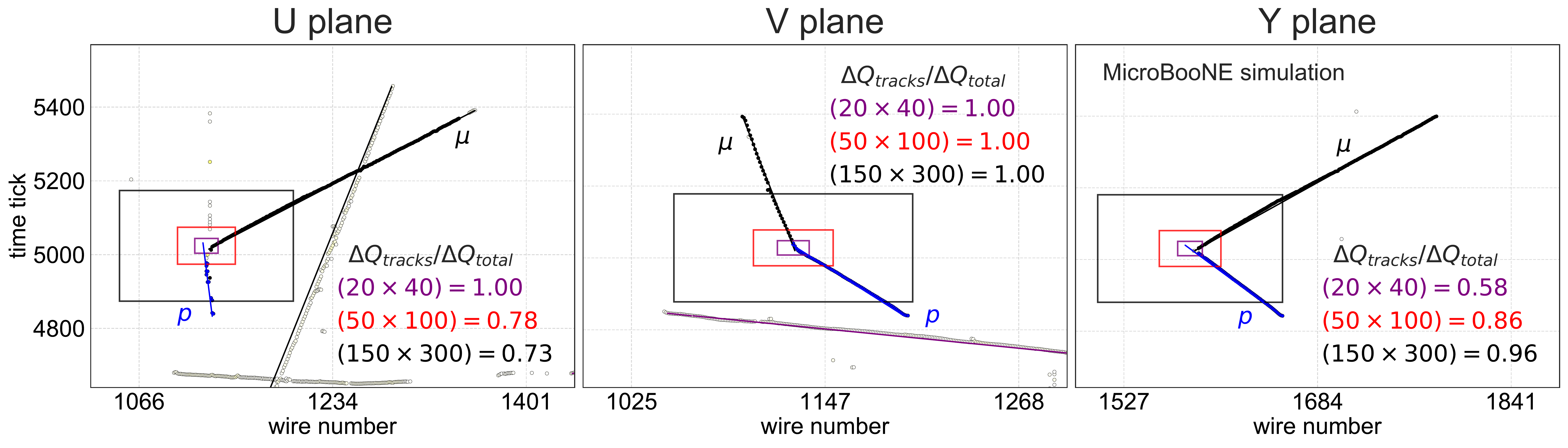}}
			\caption{A typical \CCIpOpi\ event with cosmic activity overlaid in \uB\ data.
			Shown are three boxes of different dimensions ($N_{wires}\times N_{ticks}$),
			and the ratio \RdQ\ for each,
			in the two induction planes (U and V) and in the collection plane (Y).
			In a \CCIpOpi\ event,
			\RdQ\ is expected to be close to unity for a box with dimensions that do not encapsulate too much noise
			from non--neutrino induced background.
			}
			\label{fig:CC1p0piVertex_33_vertex_2}
 		\end{figure*}

		The PandoraNu algorithm
		combines hits to form tracks,
		and associates charge deposition, $\Delta Q$, with a particle trajectory,
		allowing a measurement of the fraction of charge deposition that is not associated with reconstructed tracks in the vertex region. 
		
		For each reconstructed vertex,
		we project the position of the vertex onto each of the three wire plane views,
		and define a sequence of boxes of increasing size centered on the vertex. 
		We study the vertex activity \RdQ\ as a function of the size of the box:
		\begin{equation}
			R_{\Delta Q} = \frac{\Delta Q \left(tracks\right)}{\Delta Q \left(total\right)}.
		\end{equation}
	
		Figure \ref{fig:CC1p0piVertex_33_vertex_2} shows a typical simulated \CCIpOpi\ event in which \RdQ\ is close to unity for a box with
		dimensions ranging from 20 wires $\times$ 40 time--ticks to 150 wires $\times$ 300 time--ticks.
		The size of the box should be small enough not to capture too much irrelevant noise or background.
		
		We optimize the dimensions of the box to maximize the difference between the distribution of \CCIpOpi\ events and background.	
		Since the $R_{\Delta Q}$ objects are three-dimensional
		($\left( R_U,R_V,R_Y \right)$ where $(R_{\Delta Q})_{U} \equiv R_U$,
		and similarly for V and Y planes), 
		we need to use information from all three views to compare the distributions.
		Using the ``energy test'' \cite{Aslan:2002cn},
		we find that the optimal box dimensions are around 50 wires $\times$ 100 time-ticks, corresponding to about $15 \times 5.5$ cm$^{2}$.
		Notice that the boxes are two dimensional (wires $\times$ time-ticks) in each plane.

\begin{figure}[h]
  			\centering
		    	\includegraphics[width=\linewidth]{{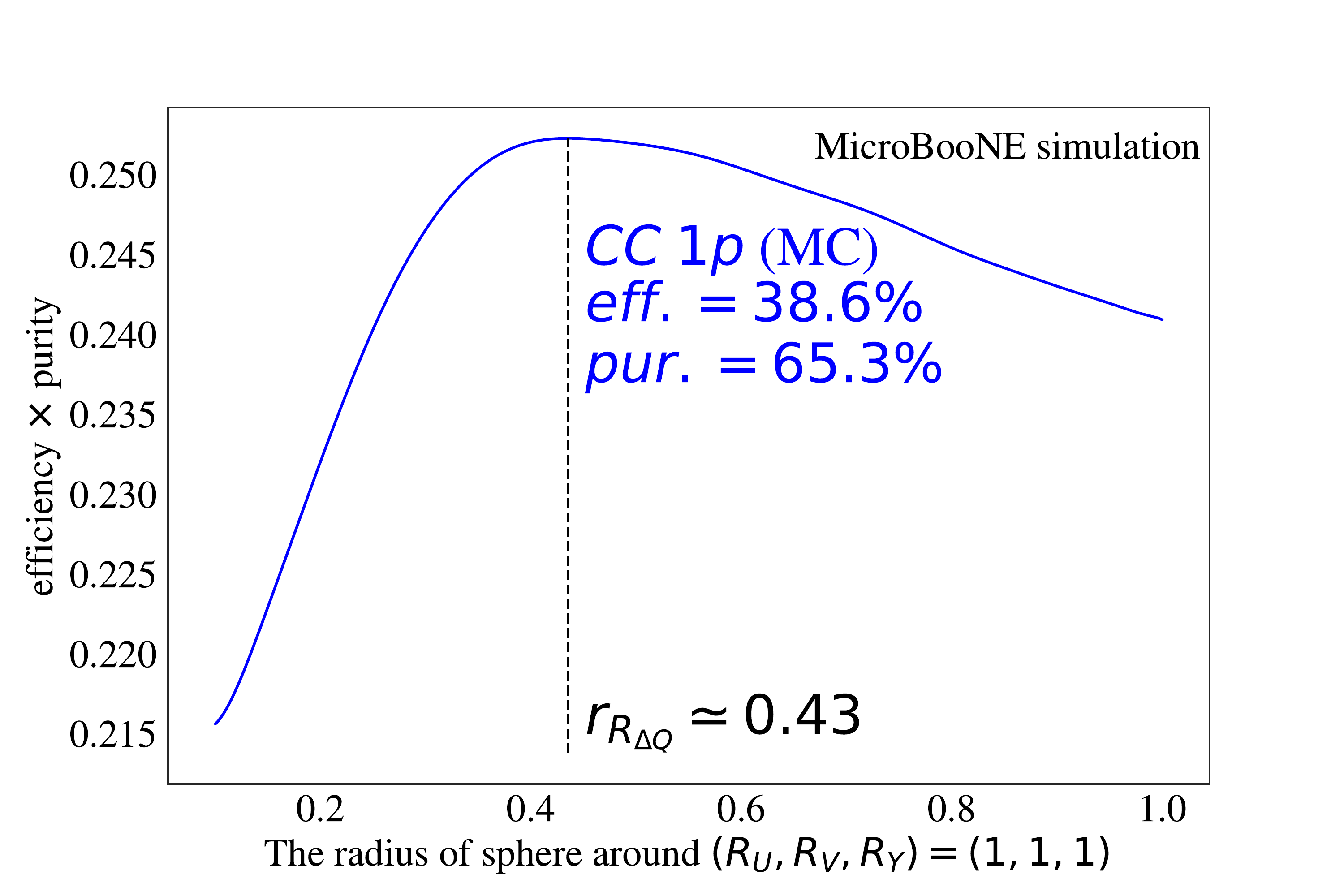}} 
			\caption{The product of the purity times efficiency for \CCIpOpi\ events as a function of $r_{R_{\Delta Q}}$,
			given by Eq. \ref{eq:RdQ cut definition}.
			The dashed line indicates the chosen value of $r_{R_{\Delta Q}}$.
			See text for details.}
			\label{fig:cut_optimization_rRdQ}
 		\end{figure}

		For simplicity, 
		we optimize a one-parameter selection in the space of the three ratios,
		the radius $r_{R_{\Delta Q}}$,
		of a sphere around $\left( R_U,R_V,R_Y \right) = \left( 1,1,1\right)$
		where $R_{U (V) [Y]}$ is $R_{\Delta Q}$ in U(V)[Y] plane:
		\begin{equation}
			\label{eq:RdQ cut definition}
			\sqrt{ \left( R_U -1 \right)^2 + \left( R_V -1 \right)^2 + \left( R_Y -1 \right)^2 } \le  r_{R_{\Delta Q}}.
		\end{equation}
		To optimize the cut on $r_{R_{\Delta Q}}$,
		we define a figure of merit equal to the product of the \CCIpOpi\ purity ($p$) and efficiency ($\epsilon$),
		\begin{equation}
			\label{eq:pur_eff}
			p \times \epsilon =
			 \left(\frac{N_{\textrm{CC1p0}\pi }^{\textrm{after cuts} } } {N_{\textrm{total}}^{\textrm{after cuts}} }\right)
			 \left(\frac{N_{\textrm{CC1p0}\pi }^{\textrm{after cuts} } } {N_{\textrm{CC1p0}\pi }^{\textrm{before cuts}} } \right).
		\end{equation}
		Figure \ref{fig:cut_optimization_rRdQ} shows the figure of merit defined in Eq. \ref{eq:pur_eff} for the \CCIpOpi\ events as a function of $r_{R_{\Delta Q}}$.
		As can be seen in Fig. \ref{fig:cut_optimization_rRdQ},
		the product of purity and efficiency is maximal for 
		$r_{R_{\Delta Q}} \approx 0.43$
		and thus we choose
		$r_{R_{\Delta Q}} \le 0.43$
		for the selection of \CCIpOpi\ events.
		
	\subsection{Coplanarity requirement} 
	
	For each track we define $\phi$ as the azimuthal angle in the $x-y$ plane,
	and the azimuthal difference between the two tracks as $\Delta \phi = \phi_{p} - \phi_{\mu}$.
	
\begin{figure}[htb!]
	    \centering	    
	    \subfigure
			{\label{fig:DeltaPhi_a}
			\begin{tikzpicture}
	        		\node[anchor=south west,inner sep=0] (image) at (0,0) {\includegraphics[width=\linewidth]{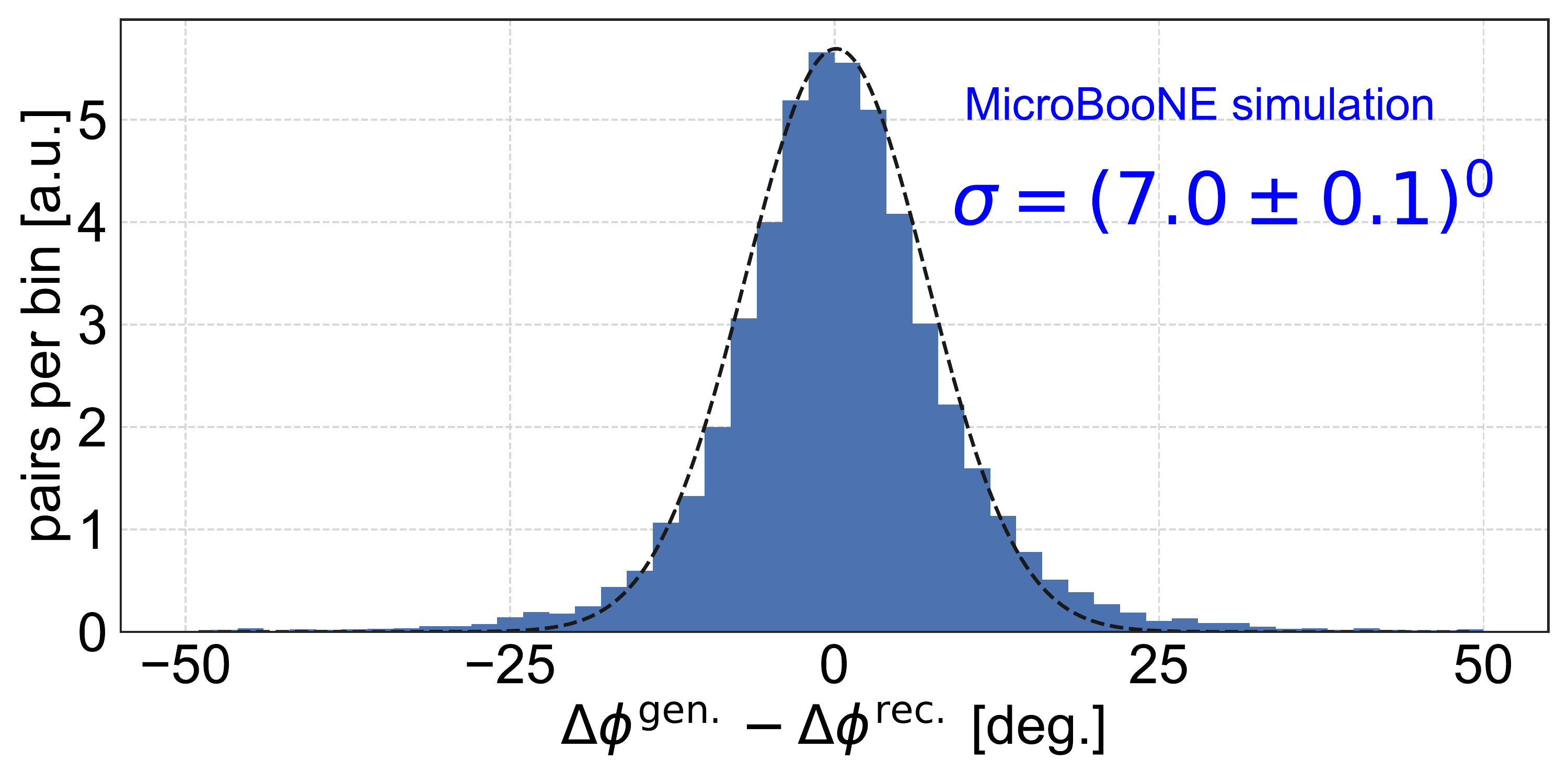}};
	        		\begin{scope}[x={(image.south east)},y={(image.north west)}]
	            		\node[anchor=south west,inner sep=0] (image) at (0.09,0.725) {\includegraphics[trim={0 17cm 16cm 0},clip,width=0.4\linewidth]{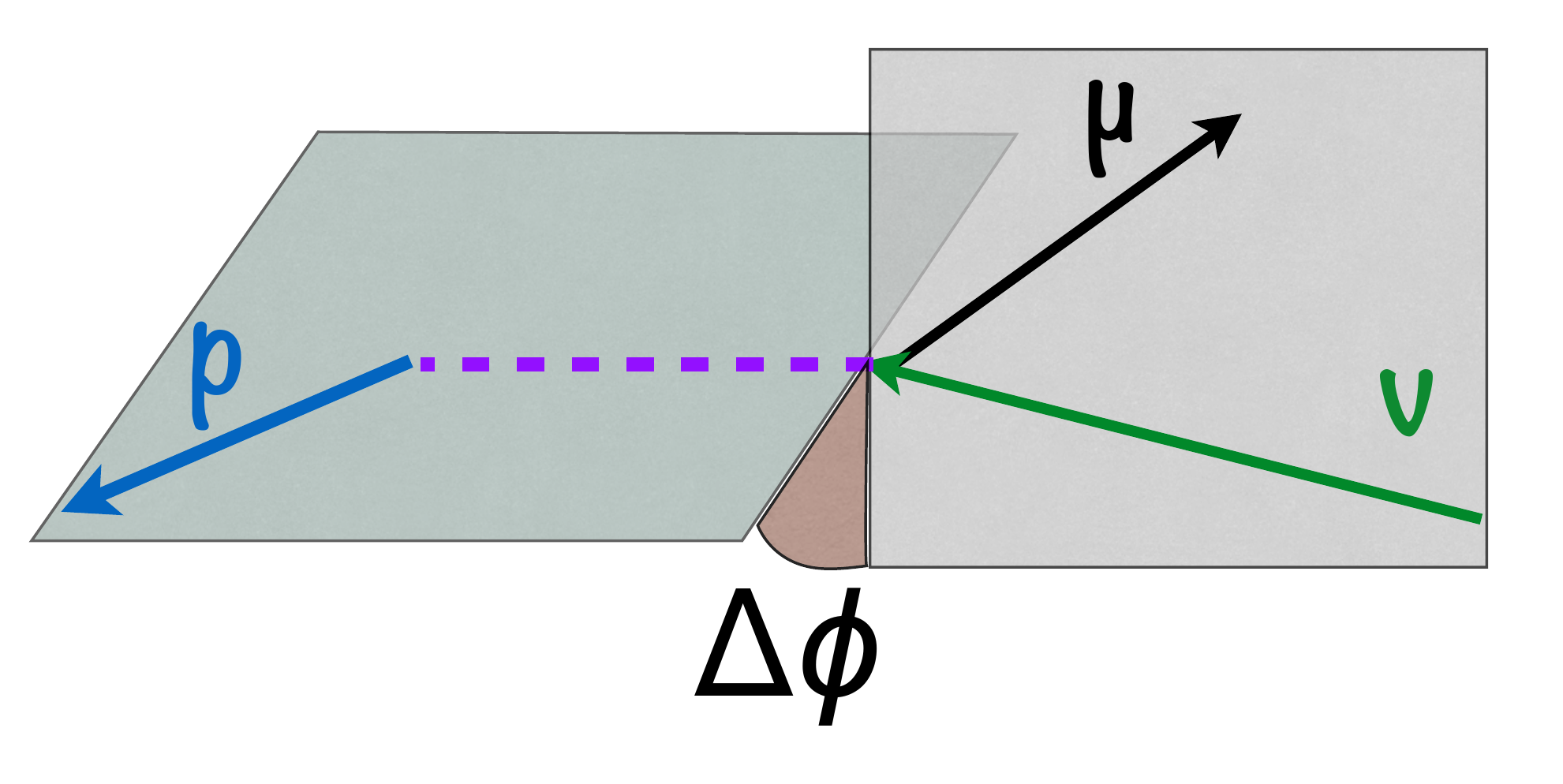}};
        			\end{scope}
	        		\begin{scope}[x={(image.north west)},y={(image.south east)}]
	            		\node[anchor=south east,inner sep=0] (image) at (0.05,0.3) {(a)};
        			\end{scope}
	    		\end{tikzpicture}
			}	    
	    	\begin{tabular}{@{}p{\linewidth}@{\quad}p{\linewidth}@{}}
		 \subfigimg[width=\linewidth]{(b)}{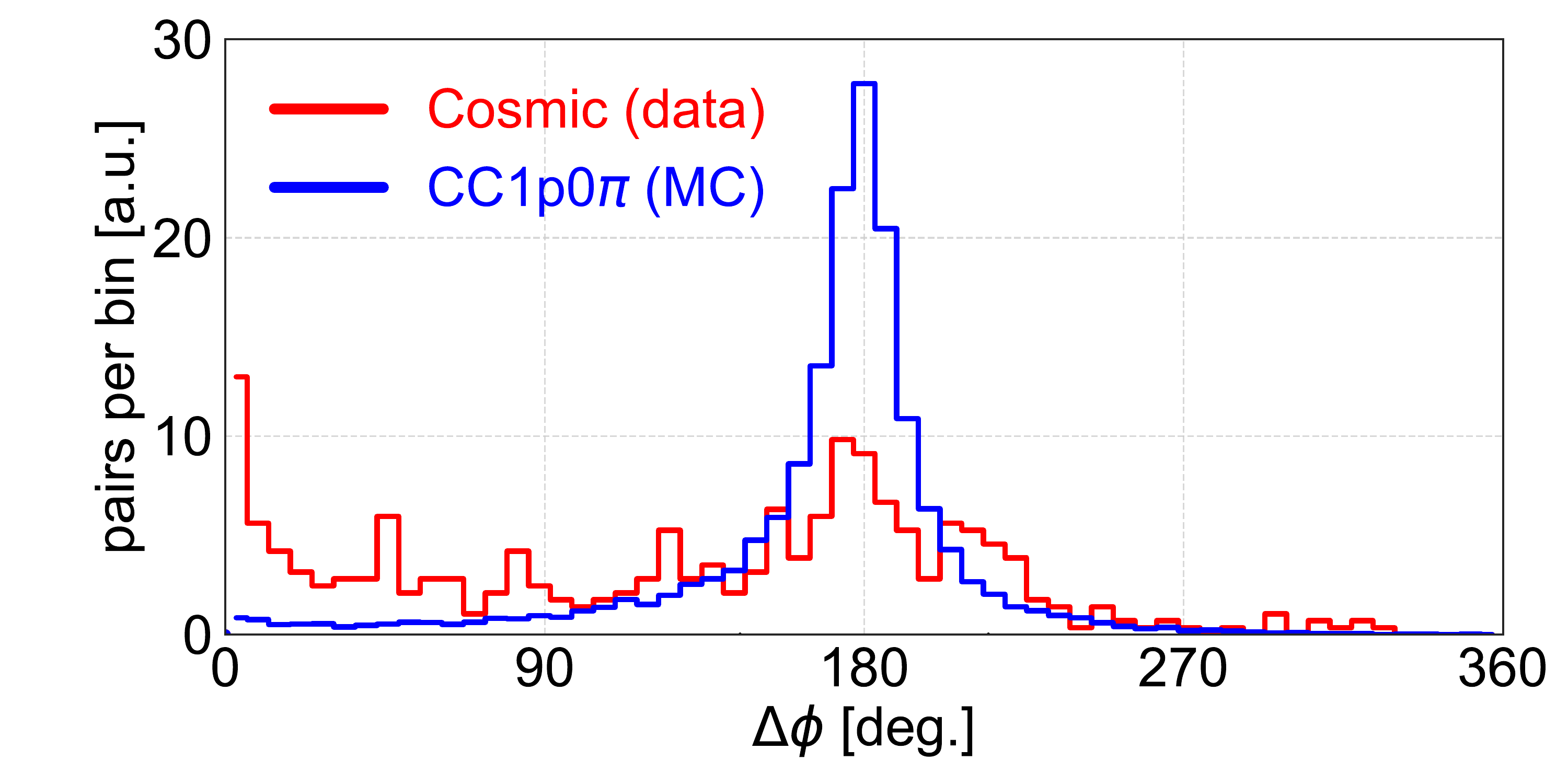}{1.7}{220} \\
		 \subfigimg[width=\linewidth]{(c)}{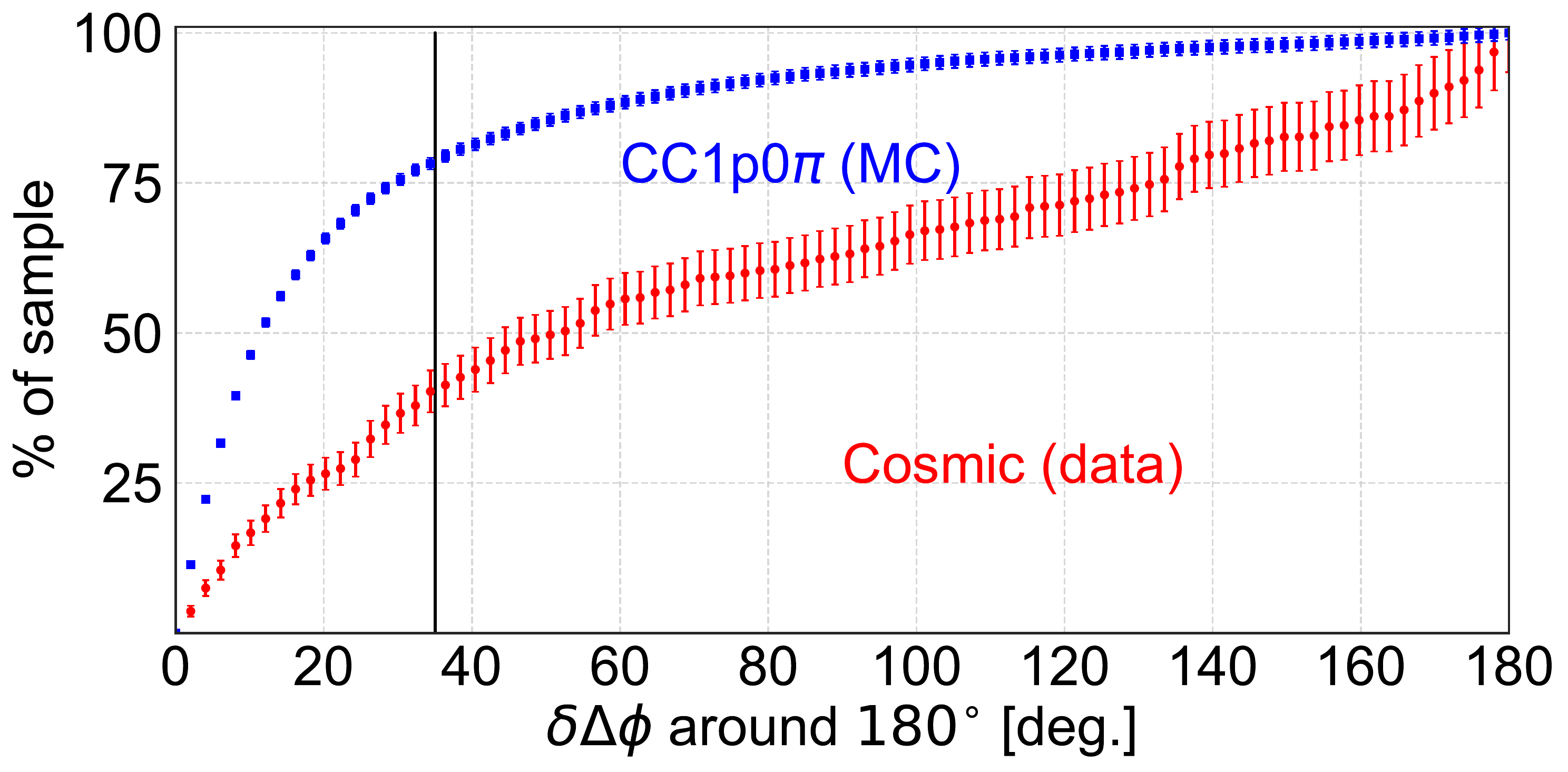}{4.4}{220} \\
		\end{tabular}	    
	    \caption{(a) The difference between the reconstructed and generated (truth level) $\Delta \phi$ for \CCIpOpi\ events in \uB\ simulation.
			A fit for a Gaussian distribution function around the peak is shown in the figure,
			as well as the width $\sigma$ of the best-fit result.
			The illustration in the insert shows the definition of the angle.
			(b)  The distributions of the reconstructed $\Delta \phi$ between the 
			candidate $\mu$ and $p$ candidates after all previous criteria were applied.
			(c) The effect of a symmetric selection around $\Delta \phi = 180^{\circ}$ as a function of the selection criterion.}
		\label{fig:DeltaPhi}
	\end{figure}

	Assuming two-body kinematics,
	the muon and proton tracks should lie on a mutual plane with the beam axis direction ($z$),
	i.e., the azimuthal angular difference between the outgoing tracks is expected to be $180^{\circ}$,
	as illustrated in the insert of Fig. \ref{fig:DeltaPhi}.
	In CCQE-like events the Fermi motion of the nucleon, nuclear re-scattering,
	and the resolution of the detector produce small deviations around $180^{\circ}$.
	Hence,
	a requirement of a large $\Delta \phi$ between the tracks suppresses contributions from events with multi--hadron production.
	To utilize the coplanarity,
	we use the reconstructed start and end points of the candidate muon and proton tracks,
	and correct the directionality of the tracks to exit from the reconstructed vertex.
	
	Figure \ref{fig:DeltaPhi}\textcolor{blue}{(a)} shows the difference between the reconstructed and generated $\Delta \phi$
	for simulated \CCIpOpi\ events after the application of all the selection criteria discussed above.
	The standard deviation of the distribution is depicted in the figure,
	and serves as an estimate of the detector resolution for $\Delta \phi$.
		
	The prominence of coplanarity in \neumu\ CCQE-like events was used
	in an analysis of similar events by the \MINERvA\ collaboration \cite{Betancourt:2017uso}.
	We find that the \uB\ coplanarity resolution is about $7^{\circ}$
	(see Fig. \ref{fig:DeltaPhi}\textcolor{blue}{(a)}),
	about twice as large as the \MINERvA\ reported value of $3.8^{\circ}$ but still sufficient for our purposes.
	This difference is primarily due to the difference in typical event kinematics,
	which results from the lower energies of \uB\ as compared to \MINERvA.

	Figure \ref{fig:DeltaPhi}\textcolor{blue}{(b)} shows the distributions of the reconstructed $\Delta \phi$ between the two tracks,
	after we apply the detection--based selection criteria described in Sec. \ref{sec:DetectorObservables}. 
	Figure \ref{fig:DeltaPhi}\textcolor{blue}{(c)} shows the effect of imposing a requirement around $\Delta \phi = 180^{\circ}$ on the different samples.
	To enhance the contribution from \CCIpOpi\ and suppress background,
	we require \CutDeltaPhi.

	\subsection{Transverse momentum imbalance}
	
	For \CCQEp\ CCQE scattering off a single neutron, with no nuclear correlations,
	the component of the total reconstructed momentum transverse to the incoming neutrino,
	\begin{equation}
		p_T = (\vec{p}_{\mu}+\vec{p}_{p})_{T},
	\end{equation}
	should be small and mainly due to the Fermi motion of the knocked out neutron,
	final state interactions of the emerging proton,
	and momentum reconstruction resolution.
	
\begin{figure}[htb!]
  		\centering
		\begin{tabular}{@{}p{\linewidth}@{\quad}p{\linewidth}@{}}
		 \subfigimg[width=\linewidth]{(a)}{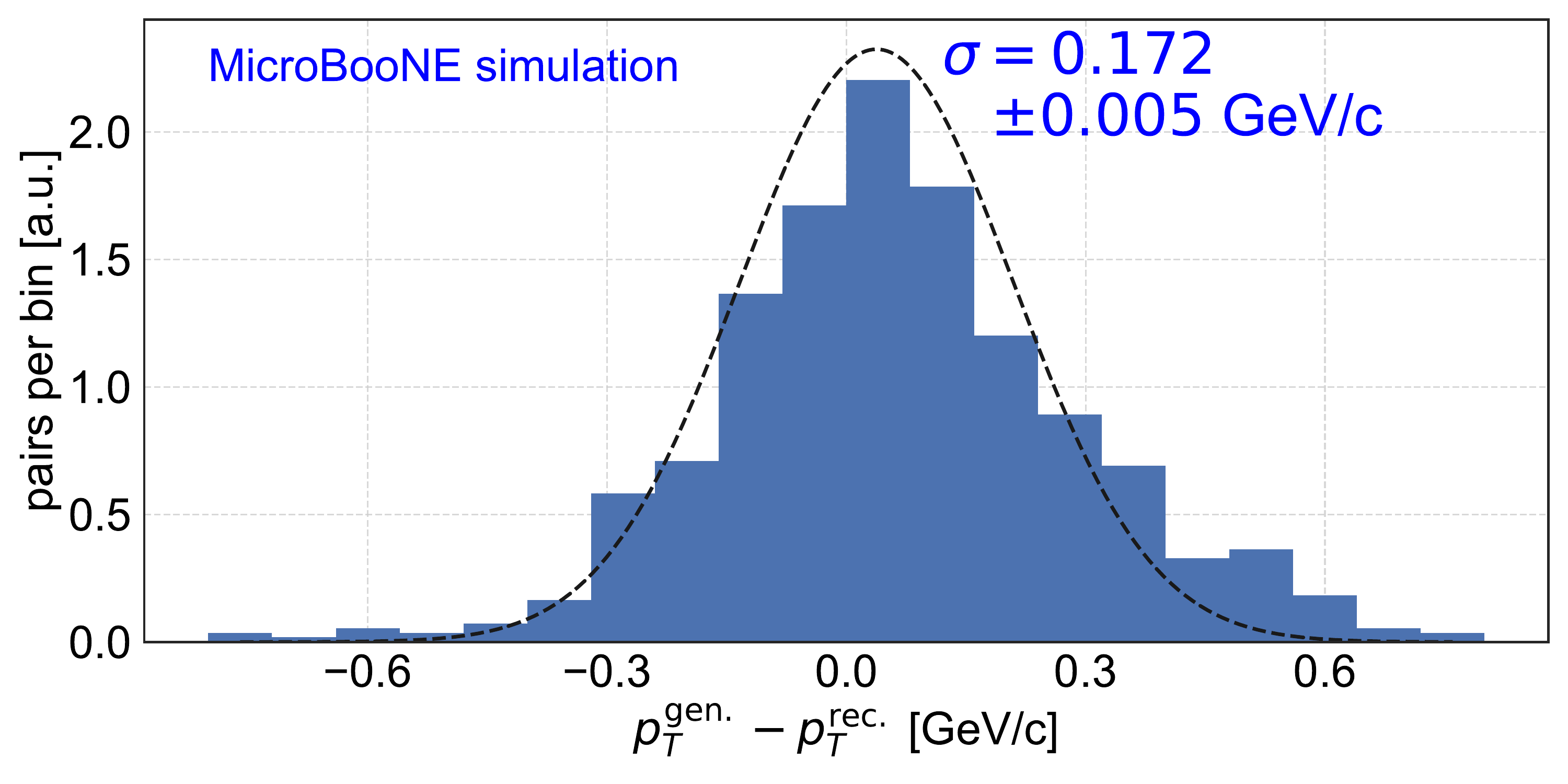}{1.3}{224} \\
		 \subfigimg[width=\linewidth]{(b)}{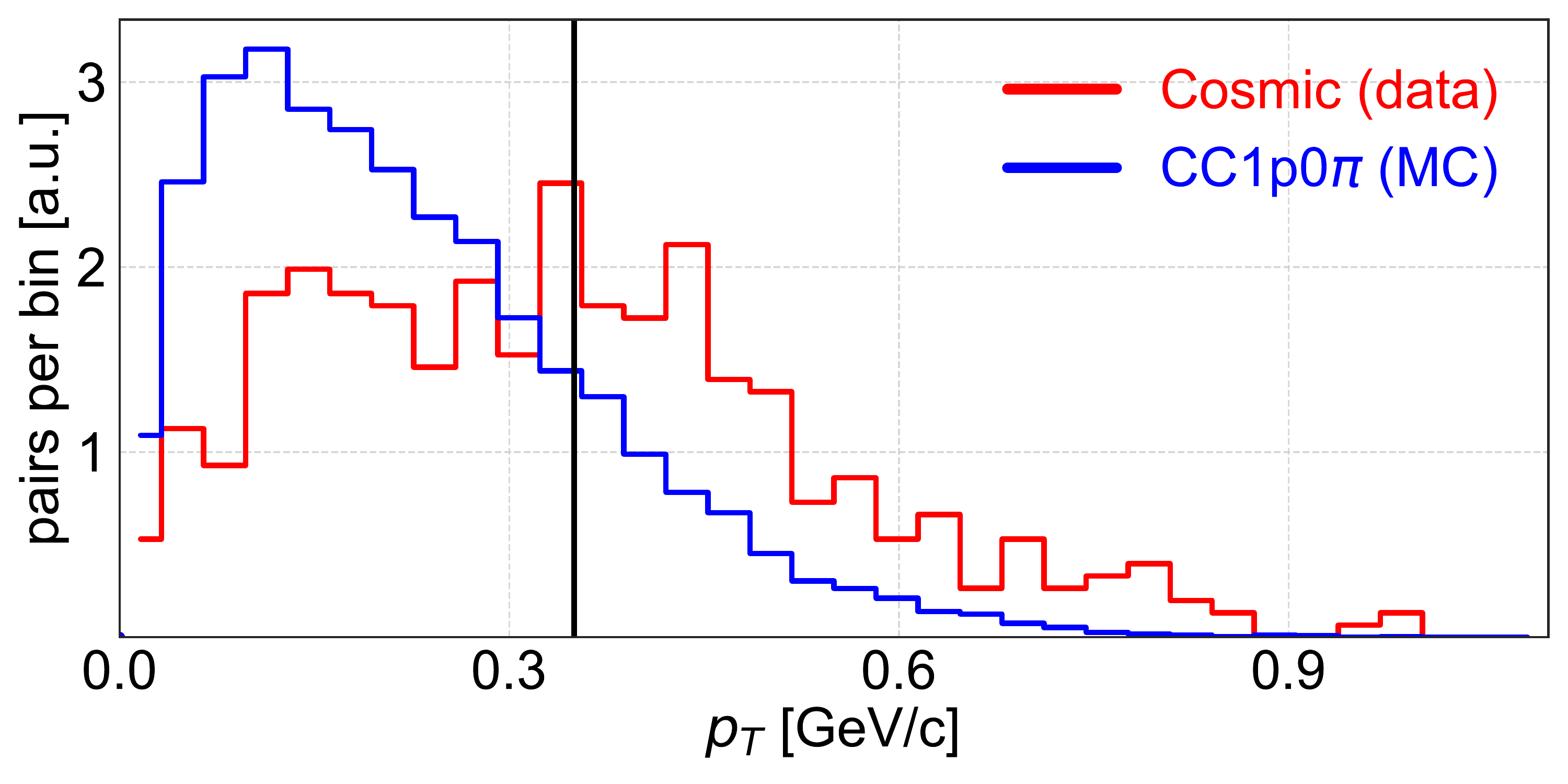}{4.2}{220} \\
		 \subfigimg[width=\linewidth]{(c)}{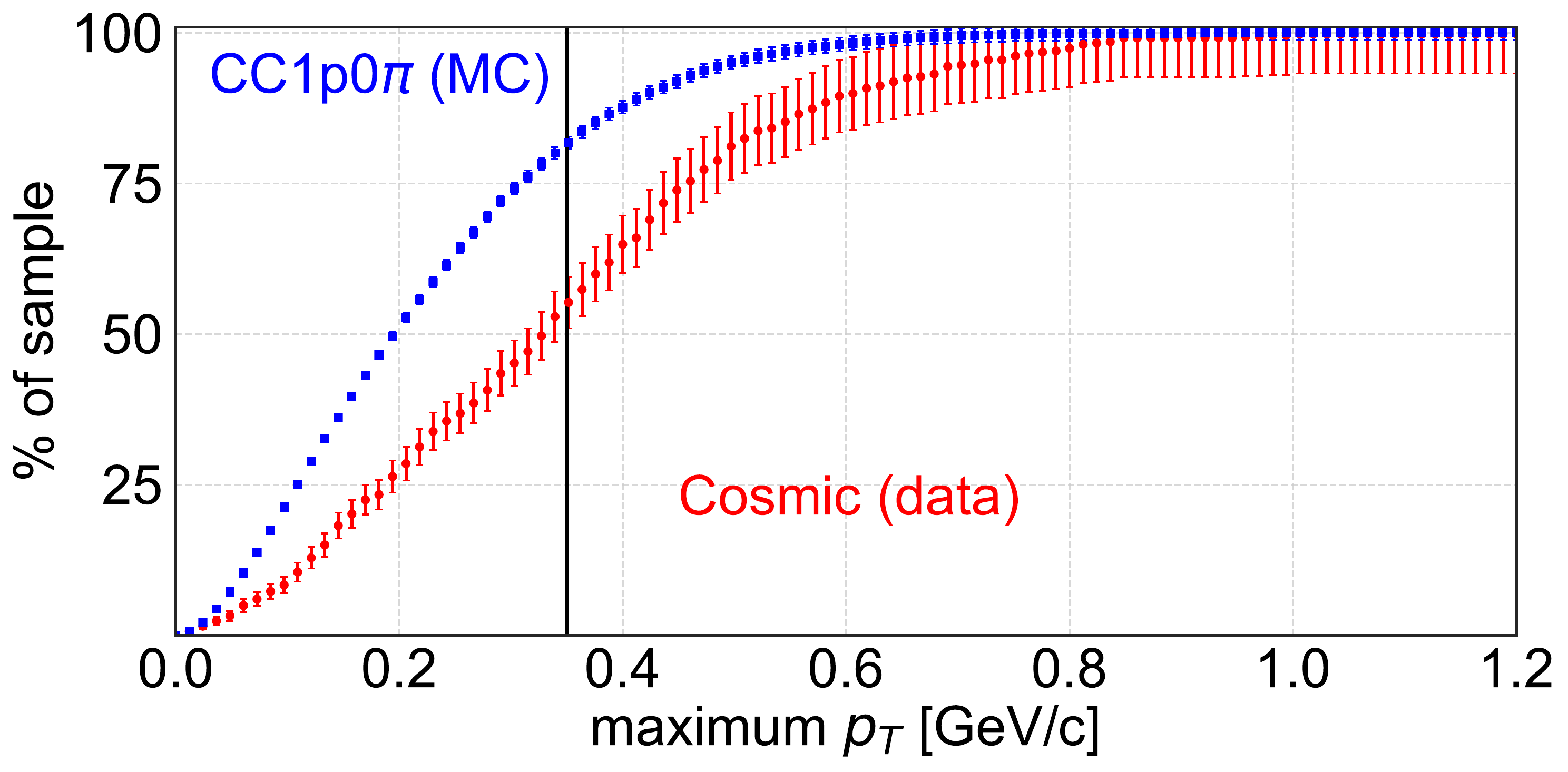}{4.4}{220} \\
		\end{tabular}	    
		\caption{(a) The difference between the reconstructed and generated (truth-level) $p_T$,
			     	for simulated \CCIpOpi\ signal events.
			     	A fit for a Gaussian distribution function around the peak is depicted in the figure,
			     	as well as the width $\sigma$ of the best-fit result.
			  	(b) The distributions of the reconstructed $p_T$ between the two tracks,
			     	after the application of the detection selection criteria.
				(c) The effect of a selection criterion on the maximal reconstructed $p_T$.}
		\label{fig:Pt}
 	\end{figure}

	We estimate the momenta of the final state particles from the stopping range of the tracks in liquid argon.	
	Figure \ref{fig:Pt}\textcolor{blue}{(a)}
	shows the difference between the reconstructed and the generated $p_T$ for \CCIpOpi\ events,
	after application of all selection criteria.
	The standard deviation of the distribution is given in the figure.
	This $\sigma$ is only used to verify that our chosen cuts are far from the measurement resolution.
	Figure \ref{fig:Pt}\textcolor{blue}{(b)} shows the distributions of the reconstructed $p_T$,
	after applying all previous selection criteria.	
	Figure \ref{fig:Pt}\textcolor{blue}{(c)} shows the effect of imposing a maximum $p_T$ on the different samples.
	To enhance the contribution from \CCIpOpi\ and suppress background,
	we require that \CutPt.



	\section{Cosmic rejection summary}
	\label{sec:Cosmic rejection with the above criteria}


	Table \ref{tab:ApplicationOfCuts} shows the sequential impact of each of the applied selection criteria discussed above.
	The original number of pairs in the simulated-signal and data-background samples, 
	labeled as ``preselection'',
	includes all events that survive CR rejection
	in \uB\ as discussed in Sec. \ref{sec:Cosmic rejection prior to this analysis}.
	Applying the detector--based requirements described above we retain $45.1\%$ of the \CCIpOpi\ simulated signal,
	while suppressing about $99.5\%$ of the CR background.
	After further selection based on the kinematics of CCQE-like interactions,
	including $\Delta \phi$
	and the reconstructed $p_T$,
	the CR background is reduced to about $0.07\%$
	of the number of cosmic events in the original sample.

	The purity of the \CCIpOpi\ simulated signal,
	listed in Table \ref{tab:ApplicationOfCuts},
	is $78.4\%$ for the 1:1 cosmic overlay assumption, and is computed by $N_{CC1p0\pi}/(N_{cosmic} + N_{beam})$,
	where $N_{beam} = 12,676$ is the number of close--track pairs induced by the simulated neutrino interactions,
	of which $78.4\%$ ($N_{CC1p0\pi}=10,020$, see last row of Table \ref{tab:ApplicationOfCuts})
	are contributed by the \CCIpOpi\ after application of the event--section requirements,
	and $N_{cosmic} = 104$.
	As discussed in Sec. \ref{sec:Cosmic rejection prior to this analysis},
	for the real \uB\ case the cosmic contribution is about 10 times larger after the application of the software trigger.
	Consequently,
	the final purity would change from about $78.4\%$ to $N_{CC1p0\pi}/( 10 \times N_{cosmic} + N_{beam}) \approx 73\%$.

	\subsection{Sensitivity to the selection criteria parameters}
	The results shown in Table \ref{tab:ApplicationOfCuts} were obtained using specific cut values.
	As part of cut optimization, we vary each of the cut parameters based on an arbitrary $10\%$ variation of the parameter,
	or its resolution estimated from simulation, whichever is larger.
	To study the variation of the final efficiency and purity,
	an ensemble of 1000 combinations of cut parameters was generated,
	each parameter chosen at random from Gaussian distributions with the following means ($\mu$) and standard deviations ($\sigma$):
			\begin{enumerate}
			\item $(\chi^{2}_{p})^{\mu} : (\mu=80 , \sigma=10)$,
			\item $(\chi^{2}_{p})^{p} : (\mu=30 , \sigma=5)$,
			\item $N_{\textrm{PE}} : (\mu=150 , \sigma=15)$,
			\item $d_{\textrm{YZ}} : (\mu=200 \textrm{ cm} , \sigma=50 \textrm{ cm})$,			
			\item $\Delta \theta_{12} : (\mu=55^{\circ} , \sigma=5^{\circ})$,
			\item $r_{R_{\Delta Q}} : (\mu=0.43 , \sigma = 0.05) $,
			\item $\delta \Delta \phi  : (\mu=35^{\circ} , \sigma = 5^{\circ}) $,
			\item $p_T^{\textrm{max}} : (\mu=0.35 \textrm{ GeV/c}, \sigma = 0.05 \textrm{ GeV/c}) $.
			\end{enumerate}			
		The CR background rejection factor and the corresponding simulated signal efficiency were computed for each randomly sampled parameters combination.
			Then, 
			the standard deviation $\sigma$ of the resulting distributions served as a measure for the sensitivity to the selection criteria parameters values.			
			
		The resulting sensitivity of the detector-based selection criteria combination,
			is $\sigma = 0.1\%$ for the CR rejection factor and $\sigma=4.0\%$ for the signal efficiency.
			The sensitivity of the combination of all selection criteria,
			is $\sigma = 0.03\%$ for the CR rejection factor and $\sigma=3.5\%$ for the signal efficiency. 			

	\section{Summary}
		The use of state-of-the-art LArTPC detectors allows measurement of the final state characteristics of neutrino-argon interactions with unprecedented detail.
	While their use in current and future neutrino oscillation experiments
	will enable a new view into neutrino physics,
	measurements from these detectors can be limited by significant cosmogenic backgrounds.
	Rejecting such backgrounds is particularly challenging for LArTPC detectors positioned on the Earth's surface, such as those to be used in the Fermilab SBN program.					
	Using a sample of cosmic data collected by \uB\ overlaid with simulated neutrino interactions generated using GENIE, 
	we present,
	for the first time,
	methods for CR background removal in exclusive CCQE-like neutrino interactions.
	The event selection criteria remove CR backgrounds based on detector observables and the kinematics of CCQE interactions.
	The net result is a suppression of about three orders of magnitude in the CR background,
	while retaining $50-25\%$ of the simulated signal events with a signal purity of about $50-80\%$,
	depending on the application of detector level cuts or the addition of kinematical cuts.
	The choice of cuts depends on the analysis goals and efficiency--purity combination required to meet these goals.
	While our study uses cosmic data collected by \uB\ and simulated neutrino interactions generated using GENIE,
	the methods presented are generic and can be adapted to other experiments that use LArTPC detectors.
	
	For example, the cosmic rejection procedure presented here was developed with the aim of testing nuclear physics models of the most basic CCQE process in well--defined kinematics.
	For that, and similar purposes,
	given the available \uB\ statistics,
	the focus was put on achieving high purity of the selected events and the price in efficiency is tolerable.
	For other purposes,
	different purity--efficiency combinations can be obtained by adopting different combination of the cuts \cite{Acciarri:2016ryt}.
	
	Implementation of the external cosmic ray tagger in \uB\ \cite{Auger:2016tjc},
	and other hardware improvements,
	are expected to allow comparable CR rejection with looser cuts that should result in higher signal selection efficiencies and comparable purities.

	\section{Acknowledgments}	
		This document was prepared by the MicroBooNE collaboration using the resources of the Fermi National Accelerator Laboratory (Fermilab),
	a U.S. Department of Energy, Office of Science, HEP User Facility.
	Fermilab is managed by Fermi Research Alliance, LLC (FRA),
	acting under Contract No. DE-AC02-07CH11359.  MicroBooNE is supported by the
	following: the U.S. Department of Energy, Office of Science,
	Offices of High Energy Physics and Nuclear Physics;
	the U.S. National Science Foundation;
	the Swiss National Science Foundation; the Science and Technology Facilities Council of the United Kingdom;
	and The Royal Society (United Kingdom). 
	Additional support for the laser calibration system and cosmic ray tagger was provided by the Albert Einstein Center for Fundamental Physics, Bern, Switzerland.
	This work was also supported by the Israel Science Foundation.
	Erez O. Cohen would like to acknowledge the Azrieli Foundation.
	
	\begin{table}[H]
	\center
	\caption{Application of the detector observable and kinematic selection criteria on close-proximity tracks contained in the detector fiducial volume.
			The requirements in the first five rows are made on variables insensitive to the neutrino interaction model (see Sec. \ref{sec:DetectorObservables}),
			and in the four last rows are the kinematic selection criteria (see Sec. \ref{sec:KinematicalSignature}).
			The numbers in parentheses are the fractions of the original studied samples retained after the application of each selection criterion sequentially.
			The purity given in the table is the relative purity for the sample under study in this analysis,
			in which every event is forced to include a neutrino interaction.
			This purity takes into account, in addition to CR background,
					additional beam-related background simulated by GENIE (which is not \CCIpOpi,
					which is not discussed in the paper.
			}
		\label{tab:ApplicationOfCuts}
		\begin{tabular}{|l|c|c|c|}
					
			\hline
									& DATA		& \multicolumn{2}{c|}{simulated signal}	 				\tabularnewline
		\hline									
			criterion	 					& cosmic		& \CCIpOpi  		& purity 			\tabularnewline

\hline 
			\multicolumn{4}{c}{preselection} \tabularnewline 
 \hline
			preselection         & 155416($100.0\%$)         & 37228($100.0\%$)         & $13.1\%$         \tabularnewline
\hline
			\multicolumn{4}{c}{detector--response requirements} \tabularnewline 
\hline
			$dE/dx$ profile         & 8327($5.4\%$)         & 25016($67.2\%$)         & $38.8\%$         \tabularnewline
\hline
			optical filter         & 2256($1.5\%$)         & 19208($51.6\%$)         & $43.6\%$         \tabularnewline
\hline
			track lengths         & 1874($1.2\%$)         & 17623($47.3\%$)         & $46.5\%$         \tabularnewline
\hline
			collinearity         & 839($0.54\%$)         & 16796($45.1\%$)         & $50.5\%$         \tabularnewline
\hline
			\multicolumn{4}{c}{kinematical requirements} \tabularnewline 
\hline
			vertex activity         & 467($0.30\%$)         & 15034($40.4\%$)         & $62.1\%$         \tabularnewline
\hline
			coplanarity only         & 189($0.12\%$)         & 11824($31.8\%$)         & $75.2\%$         \tabularnewline
\hline
			$p_T$ imbalance only         & 256($0.16\%$)         & 12261($32.9\%$)         & $69.3\%$         \tabularnewline
\hline
			$\Delta \phi $ \&  $p_T$         & 104($0.07\%$)         & 10020($26.9\%$)         & $78.4\%$         \tabularnewline
\hline

	\end{tabular}
	\end{table}


\end{document}